\documentclass{aa}

\usepackage{graphicx}
\usepackage{subfig}
\usepackage{boxhandler}
\usepackage{appendix}
\usepackage{enumitem}
\usepackage{txfonts}
\bibpunct{(}{)}{;}{a}{}{,}
\begin{document}

   \title{The effects of the cluster environment on the galaxy mass-size relation in MACS J1206.2-0847}

   \author{U. Kuchner  
          \inst{1}
          \and B. Ziegler
          \inst{1}
          \and M. Verdugo
          \inst{1}
          \and S. Bamford
          \inst{2}
          \and B. H{\"a}u{\ss}ler
          \inst{3}
            }         
   \institute{\inst{1} Department of Astrophysics, University of Vienna,
              T\"urkenschanzstrasse 17, A-1180 Vienna,\\
              \inst{2} School of Physics \& Astronomy, The University of Nottingham, University Park, Nottingham, NG7 2RD, UK,\\
              \inst{3} European Southern Observatory, Alonso de Córdova 3107, Casilla 19001, Santiago, Chile\\
              \email{ulrike.kuchner@univie.ac.at}
             }

   \date{}

 
  \abstract
{The dense environment of galaxy clusters strongly influences the nature of galaxies.  Their abundance and diversity is imprinted on the stellar-mass--size plane.
Here,  we study the cause of the size distribution of a sample of $560$ spectroscopic members spanning a wide dynamical range down to 10$^{8.5}$M$_{\odot}$ ($\mathrm{log(M})$-2) in the massive CLASH cluster MACSJ1206.2-0847 at $z=0.44$. 
We use Subaru SuprimeCam imaging covering the highest-density core out to the infall regions (3 virial radii) to look for cluster-specific effects on a global scale. We also compare our measurements to a compatible large field study in order to span extreme environmental densities.

This paper presents the trends we identified for cluster galaxies divided by their colors into star-forming and quiescent galaxies and into distinct morphological types (using S\'ersic index and bulge/disk decompositions).
We observed larger sizes for early type and smaller sizes for massive late type galaxies in clusters in comparison to the field. We attribute this to longer quenching timescales of more massive galaxies in the cluster.
Our analysis further revealed an increasing importance of recently quenched transition objects (``red disks''), where the correspondence between galaxy morphology and color is out of sync.
This is a virialized population with sizes similar to the quiescent, spheroid-dominated population of the cluster center, but with disks still in-tact, and found at higher cluster-centric radii.
The mass-size relation of cluster galaxies may therefore be understood as the consequence of a mix of progenitors formed at different quenching epochs.
We also investigate the stellar-mass  -- size relation as a representation of galaxy sizes smoothly decreasing as a function of bulge fraction. 
We find that at same bulge-to-total ratio and same stellar mass, quiescent galaxies are smaller than star-forming galaxies. This is likely because of a fading of the outskirts of the disk, which we saw in comparing sizes of their disk-components. Ram-pressure stripping of the cold gas and other forms of more gradual gas starvation are likely responsible for this observation. 
 }


   \maketitle
%

\section{Introduction}
One way to independently probe the evolutionary state of galaxies is through classical scaling relations that link shapes and physical sizes of the stellar distribution with galaxy luminosity \citep{kormendy77} and stellar masses \citep{shen03}.
Both the average size of a galaxy at a given stellar mass and stellar population distribution hold information of the assembly history of a galaxy.
The stellar-mass -- size relation (MSR) shows distinct trends for spheroid and disk-like morphologies \citep{kauffmann03}.
Independent of whether galaxies are classified on the basis of star-formation activity or structure, early type galaxies fundamentally follow a steeper dependency between size and stellar mass than late type galaxies do \citep{lange15}. 
That means that at a given mass (at least below 10$^{11} \mathrm{M}_{\odot}$) and redshift, early-type galaxies are generally smaller (as measured by the half-light radius r$_e$) than late type galaxies. 

Over the past few decades, measurements of the MSR from high-resolution observations of high redshift galaxies have lead to the important discovery that massive galaxies have experienced a dramatic evolution in size, growing at least by a factor of four at fixed stellar mass since z$\sim$2 \citep[e.g.,][]{navarro00, daddi05, trujillo06, vanderwel08}.
Current models of galaxy evolution favor two main mechanisms to account for this observed size growth. The first category ascribes internally driven in-situ processes connected to the mass of the galaxy, i.e., AGN activity \citep{croton06, fan08, fan10, ragone11} or SN-winds \citep{damjanov09} to account for mass loss and subsequent expansion of galaxies that react to the change of the gravitational potential. 
An alternative explanation, favored by a growing number of authors, explains the size growth of massive galaxies through major merging \citep{ciotti01,naab07,nipoti10} or repeated dry minor mergers \citep{khochfar06,maller06, hopkins09a,naab09,sommer-larsen10,oser10}. 
Recent observations seem to favor this scenario suggesting that high mass galaxies grow in an inside-out fashion, i.e., the cores of the galaxies have been in place at high redshifts and the evolution is seen in the outer parts \citep{vandokkum10, patel12, perez13}, thereby increasing their size 
and stellar mass continuously \citep[e.g.,][]{vandokkum10,trujillo07,bezanson09}.

The physical evolution of galaxies also depends on the environment they reside in. 
More specifically, dense environments are believed to accelerate galaxy evolution, producing more massive systems faster \citep[e.g.,][]{delucia04,gao05, maulbetsch07,shankar13}.
The direct role of the cluster environment in driving galaxy evolution through cluster specific effects, however, is still largely unclear. 
The results from extensive field surveys provide an intriguing picture of the general size evolution with cosmic time, but understanding how the environment effects the evolution of the MSR is much less definite. 
Publications from a variety of data sets and across a broad range of epochs, mostly confined to massive systems owing to our inability to detect or unambiguously measure low surface brightness galaxies at higher redshifts, report contradicting findings. 
Several authors have found that at higher redshifts (1<z<2), ellipticals have larger sizes when they belong to groups or clusters \citep{cooper12, papovich12, delaye14}. 
This might be confined to lower mass galaxies only, since, for the same redshift range, other observational studies do not find any differences for high mass galaxies \citep{rettura10}.
The controversy does not end for low-redshift observations. \citet {huertas13} report no significant environmental dependence for high mass elliptical galaxies in the local universe and postulate that the mass-size relation of early type galaxies at z$\sim$ 0 is independent of the mass of the host halo.
Conversely, \citet{raichoor12} and \citet{poggianti13} have found that ellipticals are more compact in clusters. This effect might be redshift dependent \citep{lani13}.
For spiral galaxies, observations continue to identify a trend for smaller galaxies in dense regions in comparison to the field \citep{maltby10,fernandez-lorenzo13, cebrian14}.

Making sense of the debate around the stellar-mass -- size relation is a complex challenge since, for example, alternative classification systems and the size dependence on wavelength add unwanted biases. 
The intricate nature of galaxy evolution in clusters adds a further complexity to the interpretation.
While the variations of the star-forming galaxy stellar mass function of clusters with global environment are found to be small and subtle, at least above $10^{10.25} M_{\odot}$ \citep{annunziatella14}, processes inside clusters seem to enhance a morphological transformation and suppression of star-formation: at very high and very low masses, passive galaxies are generally excessively abundant in clusters.
Direct cluster-specific effects are considered responsible for a removal of the outer halo gas reservoir, resulting in changes to more compact morphologies. Depressed gas contents also lead to lower star-formation rates and redder colors \citep[e.g.,][]{treu03,poggianti06,peng14}. 

Galaxy harassment, strangulation, suppressed accretion and ram pressure are all mechanisms by which gas of the galaxy halo or disk are either pushed out or quickly used up, essentially removing the reservoir of fuel necessary to form new stars. The gas is stripped away and joins the intracluster medium while the disk of the galaxy will fade and redden as its stars age. 
Star-forming galaxies entering a dense environment like a cluster are expected to be efficiently quenched on their first infall within 1-3 Gyrs (6 Gyrs at most) at fixed stellar mass \citep[e.g.,][]{wetzel13, hirschmann14,mok14, oman16}. From comparing morphology-density- with color-density-relations as a function of mass, we know that the environment affects star-formation in low-mass galaxies more strongly than their structure \citep{bamford09}. They transform much more rapidly than high mass galaxies that quench on longer timescales.
However, these processes do not affect the galaxy uniformly \citep{bekki09}.
The gas is less strongly bound in the outskirts of the disk, so will be removed there first. As a result, the mass-to-light ratio changes rapidly with radius, and the star-formation rate drops farther out. The stripped disk will fade with time, thus eventually changing the stellar mass profile.
In addition, tidal stripping, either through the cluster potential or interactions with other galaxies, may be removing outer parts of the stellar and gaseous disks \citep{boselli06}.

\bigbreak
To form a better view of the global evolution of the stellar-mass -- size relation of galaxies in clusters, we have studied the size distribution of cluster members, its dependence on stellar mass, morphological type, star-forming class and environment. 
We pay particular attention to differences that arise from separating galaxy populations in a number of ways, and use the broader terms early/late type galaxies only to describe the general bimodality of galaxies.

For this investigation, we use deep multiband images from Subaru of the cluster MACS J1206.2-0847, part of the comprehensive data set of the Cluster Lensing and Supernova Survey with Hubble \citep[CLASH, ][]{postman12} and CLASH-VLT \citep{rosati14}. The combined data allow a detailed study of 560 spectroscopically confirmed cluster galaxies, divided into different types, according to:

(i) Colors: We used the colors of the cluster galaxies to differentiate between star-forming and quiescent galaxies.

(ii) S\'ersic indices: We applied the widely used division of n=2.5 for a rough separation into disk-dominated galaxies (n<2.5) and spheroid-dominated galaxies (n>2.5).

(ii) Bulge-disk decomposition: This enabled us to further investigate the galaxies according to their bulge-to-total ratio. 

For the first time, we provide mass-size relations inside a massive cluster out to three virial radii using information from bulge-disk decomposition.
This was made possible through the high level of quality and richness of the CLASH, CLASH-VLT and complementary ground based Subaru data that allowed us to include a census of bulge and disk measurements as additional valuable information to the exploration of the MSR.   
To attribute any observed differences of the stellar-mass -- size relations in clusters from the field to a physical evolution, we need to ensure accurate measurements of the spatial structure of the galaxies. In this study, we did this by measuring all structural parameters (sizes, derived from half-light radii $r_e$ of the 2D S\'ersic profile fitting, S\'ersic indices, and fluxes for bulges and disks) in a consistent manner using five bands from Subaru Suprime-Cam (B,V,R,I,z) simultaneously, employing the multi-band fitting code GALAPAGOS-2 developed by the MegaMorph project \citep{bamford11} that uses an extension of the widely used image analysis algorithm GALFIT.
In addition to our investigation of the global properties of galaxies, we also considered the bulges and disks as independent components to correlate the sizes of the disk components of galaxies to their star-formation status. 

\bigbreak
This paper starts by explaining the sample (Section \ref{sec:data}) and measurements (Section \ref{sec:measurements}) of structural parameters and stellar masses. 
We go on to look at the stellar-mass -- size relation in the cluster in comparison to field measurements (Section \ref{sec:MSR_field_cluster}) and dissect variations inside the cluster that stem from different classifications and their combinations (Section \ref{sec:comp}). 
These considerations lead to an investigation of transitional objects, like disk-dominated quiescent galaxies (``red disks'') that play a prominent role in galaxy clusters. We then use our two-component measurements to investigate the MSR as a function of bulge fraction (Section \ref{sec: MSR_hubble_all}) and find further answers in our comparison of the disk component of star-forming and quiescent galaxies (Section \ref{sec:MSR_disk}). 
We then look at the spatial distribution of galaxies inside the cluster and introduce the cluster demographics of the galaxy population (Section \ref{sec:fractions}).
Finally, we provide an in-depth look at the MSR at different regions of MACS1206 (Section \ref{sec:MSR_radial bins}) and offer a discussion that ties our findings to possible galaxy transformation scenarios inside the cluster (Section \ref{sec:discussion}).
Throughout this paper we assume a $\Lambda \mathrm{CDM}$ 
cosmology with $\Omega_{0}=0.3$, $\Omega_{\Lambda}=0.7$, and $H_0 = 70 \mathrm{km\ s}^{-1}$ and AB magnitudes. These parameters give us a physical scale of 5.68 kpc/arcsec at z=0.44, the redshift of this cluster.

\section{Analysis}

\subsection{Data and Sample Selection}
\label{sec:data}

Our work draws from observations obtained as part of the recently finished Cluster Lensing and Supernova survey with Hubble \citep[CLASH;][]{postman12} and CLASH-VLT  survey \citep{rosati14}, its comprehensive spectroscopic follow-up.
In this study, we especially profit from analyzing the complementary archival Subaru wide-field imaging campaign that covers the cluster in 5 optical filters (B, V, R$_c$, I$_c$, z$^{\prime}$) out to 3 virial radii \citep{umetsu12}. 
The combination of the large ($34^{\prime} \times 27^{\prime}$) field-of-view of Suprime-Cam \citep{miyazaki02} and wealth of corresponding Very Large Telescope (VLT)/VIMOS spectroscopic data \citep{rosati14} allow us to present a unique and comprehensive picture of the size distribution for 560 members of the massive cluster MACS J1206.2-0847 (hereafter MACS1206) across many orders of magnitudes in densities. The overlapping CLASH data from high-resolution HST observations in 16 filters with the Wide Field Camera 3 \citep[WFC3;][]{kimble08} and the Advanced Camera for Surveys \citep[ACS;][]{ford03}, produced following the approaches described in \citep{koekemoer11}, allow us to check the quality of our ground-based structural measurements.

MACS1206 is a well studied \citep{biviano13} X-ray luminous cluster \citep{ebeling09} at z = 0.439, known to be relatively relaxed but with a few overdensities and an elongated structure, reminiscent of past accretion \citep{girardi15}.
The larger area of Subaru Suprime-Cam multi-colour imaging is ideal for examining the effect of the environment on galaxy evolution, thanks to its image quality, field of view and depth. 
The seeing FWHM in the co-added Subaru mosaic images are $1\farcs01$ in B (2.4 ks), $0\farcs95$ in V (2.2 ks), $0\farcs78$ in R\_{c} (2.9 ks), $0\farcs71$ in I$_c$ (3.6 ks) and
$0\farcs58$ in z$^{\prime}$ (1.6 ks), with a limiting magnitude of R$_c$ = 26.2 for a 3$\sigma$ limiting detection within a $2\farcs0$ diameter aperture \citep[see also][for further descriptions of the observations and data reduction process]{umetsu12,umetsu14,nonino09}.
   \begin{figure} 
      \centering
        \includegraphics[width=\columnwidth]{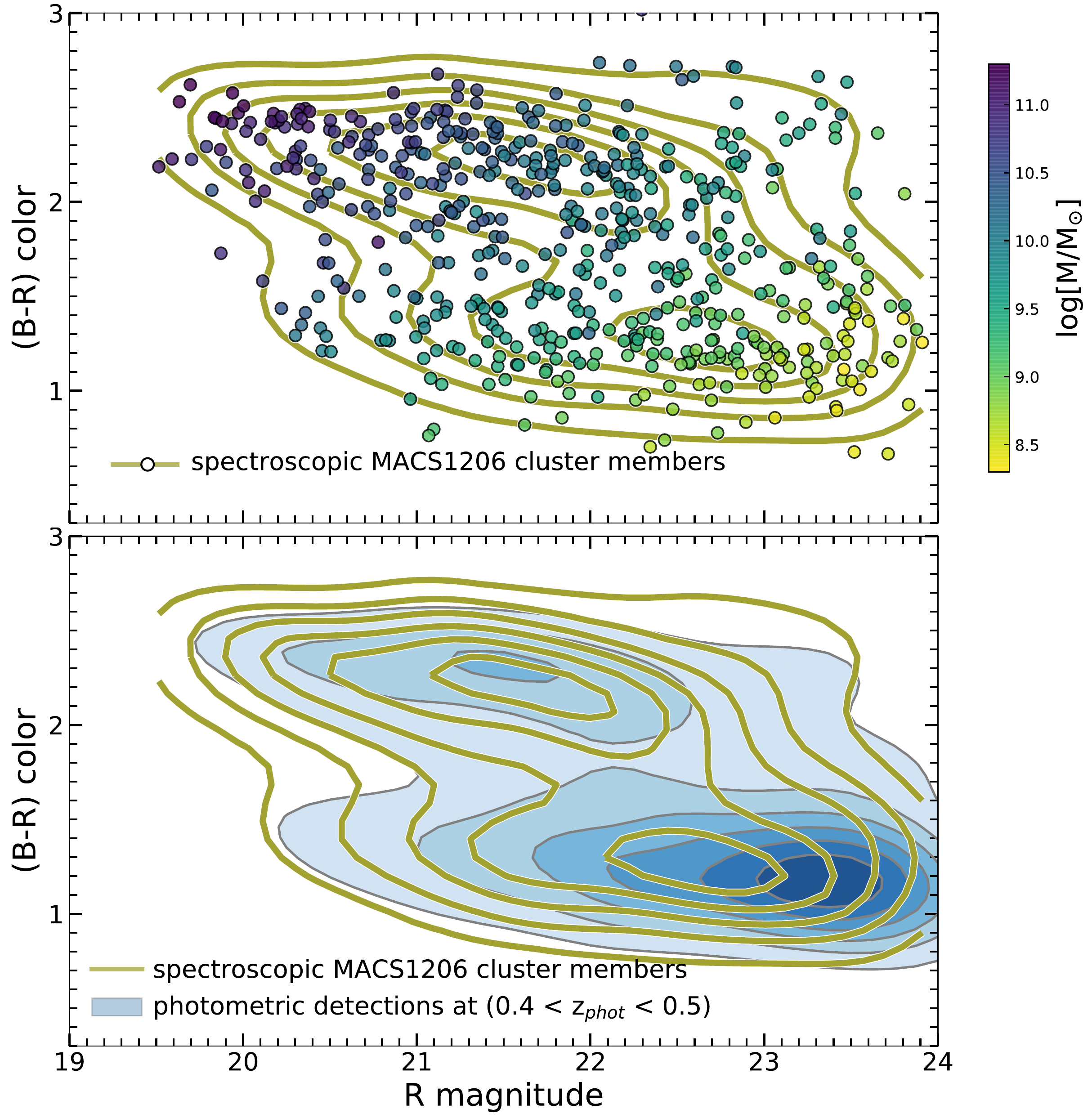}
   \caption{Observed color-magnitude relation of spectroscopic members of MACS1206. Each galaxy has been color-coded according to their mass. In the lower panel we contrast our sample (yellow density contours) to the color distribution of Subaru Suprime-Cam detections in the same field of view and at similar redshift range (blue filled density contours). This allows a rough visualization of completeness for our sample. We provide more detailed 2D completeness maps in the Appendix.
   }
        \label{fig:CMD}%
    \end{figure}

Building onto the CLASH HST panchromatic imaging project, the large spectroscopic campaign CLASH-VLT was carried out as an ESO VIMOS Large Programme \citep{rosati14}. 
It was designed to spectroscopically identify and confirm large samples of cluster members as well as lensed background galaxies for all CLASH clusters accessible for the VLT. 
VIMOS observations for MACS1206 were carried out in 2012 with a total exposure time of 10.7 hours.
To increase the exposure time for faint lensed objects behind the cluster, the cluster core was covered by a quadrant during each pointing. 
The data were reduced with the VIMOS Interactive Pipeline and Graphical Interface (VIPGI) software and given quality flags indicating the reliability of a redshift measurements. 
The final MACS1206 CLASH-VLT catalog comprises 2749 objects with reliable redshift estimates down to R$_c$ = 24.0 (see also \citet{biviano13, annunziatella14, mercurio15} for more information on slit assignment and procedure).

Fig. \ref{fig:CMD} gives an overview of the selected sample used in this study.
It shows the distribution of the B-R observed colors versus R apparent magnitudes for the final sample of 560 cluster galaxies we consider in this study. To assign cluster membership, we use spectroscopic observations of both the VIMOS low-resolution LR-blue grism and medium-resolution (MR) grism, covering an area of approximately 10 Mpc$^2$. 
Yellow contours show their distribution using a Gaussian kernel density estimation (KDE). We contrast these to the color distribution of galaxies of the underlying photometric sample (in blue) at same field of view and photometric redshift range in the bottom panel for a quick consideration of completeness.

The spectroscopic target selection was based on the photometric catalogues of the archival Subaru data and carried out in twelve masks to span an area of 20 square arcminutes. 
The difficulty of arranging slits in the core of the cluster results in a slightly higher incompleteness toward the center and for faint magnitudes. This offset is visible in the bottom panel of Fig. \ref{fig:CMD}.
In our study, we address the under-representation of faint objects by applying the inverse of the completeness as weights to the statistical properties of the cluster galaxies. 
For this, we construct completeness functions at different cluster-centric radii. In magnitude bins, we calculate the fractions of galaxies for which we successfully obtained redshifts from the total number of photometrically detected objects in the VIMOS area down to the photometric limit of $R_c = 24$.
The effect of this correction leads to increasing weights for galaxies with lower luminosities.

We further studied the on-sky completeness of our sample in relation to colors, sizes and galaxy concentration as defined by R90/R50 -- the ratio of galaxy radii containing 90\% and 50\% of the measured flux \citep[e.g.][]{blanton01, goto03, graham05} -- to test any possible bias of the spectroscopic sampling. We refer to the Appendix for these plots. While we are aware of, and correct for, the cut-off at the faint end, we do not see any biases in these other completeness fractions.

The lack of faint passive galaxies is visible both in the spectroscopic as well as in the photometric sample, which we consider to be a complete representation of the cluster, at least for R<24\,mag. This is consistent with reports of passive cluster galaxies at higher redshifts
where typical luminosity (and mass-) functions decrease at the faint end. This implies that the cluster is in the process of building its red population, and has not yet quenched all of its star-forming galaxies (see \citealt{annunziatella14} for mass-functions of MACS1206). 

For much of this paper, choices for the target selection do not have a great effect since we consider galaxies at fixed stellar mass. We do however perform the weighting for the spectroscopic incompleteness described above and consider galaxies above limits of R<24\,mag, which corresponds to $\log$ (M/M$_\odot$) = 8.5
for star-forming and R<23.5\,mag = $\log$ (M/M$_\odot$) = 9.2 for quiescent galaxies (see Appendix \ref{sec:appendix} for details.) However, we find that choosing to omit the weights does not make any significant differences to the results.
An in-depth discussion of the completeness as a function of magnitude and 2D radial distance is presented in \citet{biviano13} who explain the varying success rates of redshift measurements of bright and faint objects with differences in their signal-to-noise ratios of their spectra.

Our sample is comprised of spectroscopic cluster members, the only way to establish cluster membership with high confidence.
To distinguish between cluster members and interloping foreground and background galaxies, we assume a cluster mass model assuming a singular isothermal sphere \citep[SIS;][]{carlberg96}, thereby identifying cluster membership on the basis of their location in projected phase space. 
We use the velocity dispersion $\sigma \sim 1087 \mathrm{km/s}$, virial radius $\mathrm{R_{200}} \sim \mathrm{1.96}$, and virial mass $\mathrm{M_{200}} \sim 1.37 \times \mathrm{10^{15}} \mathrm{M_{\odot}}$ presented in \citet{biviano13} for this cluster. For our purpose, we follow the simple approach offered by \citet{carlberg97} to identify cluster members as those galaxies with velocities $|\mathrm{v}|<2\sigma(\mathrm{R})$, which is in rough correspondence with the more detailed analysis by \citet{biviano13}. 
We also use galaxies between $2\sigma(\mathrm{R})<|\mathrm{v}|<6\sigma(\mathrm{R})$, which are considered galaxies that are falling into the cluster (see Fig. \ref{fig:SIS_struc} for a visualization of the phase-space of MACS1206).
This range allows a larger sample of galaxies that have been accreted a long time ago as well as those that are part of the recent assembly history. 
Cluster-centric radii $R$ are measured relative to the Brightest Cluster Galaxy at $\alpha (\mathrm{J2000}) =
12\mathrm{^h} 06\mathrm{^m} 12\mathrm{^s}.15$, $\delta (\mathrm{J2000}) = -8\degr 48\arcmin 3\farcs4$ 
and normalized to the empirically determined $R_{200}$, defined as the radius where the mean interior overdensity is 200 $\rho_c$. As described in \citet{biviano13}, we also refer to the radius at $R_{200}$ as the ``virial radius''.

\subsection{Determination of structural parameters: sizes and morphologies}
\label{sec:measurements}

To obtain sizes for the cluster galaxies, we fit 2D Sérsic models to all cluster galaxies.
For all cluster members we measure the relevant structural parameters S\'ersic index $n$ (describing the profile shape), the half-light (i.e., effective) radius $r_e$, and (model-) luminosities both with a single S\'ersic function and for bulges and disks separately. 
The S\'ersic profile \citep{sersic68} is a characterization of the intensity $I(r)$ of the galaxy as a function of radius.
\begin{equation}
I(r)=I_e exp \left[-b_n\left(\left(\frac{r}{r_e}\right)^{1/n}-1\right)\right],
\end{equation}
where $I_e$ marks the intensity at the effective radius $r_e$ of the galaxy, which is then converted to kpc and taken as a measure of the size of the galaxy. 

For this single+multi-component morphology fitting, we exploit the capabilities of the galaxy profile model fitting software GALAPAGOS-2 developed within the MegaMorph project \citep{bamford11, haussler13}. 
This analysis tool is based on GALFITM, a multi-band extension of GALFIT3 \citep{pengC10} which enables us to utilize all information available over the five Subaru BVR$_c$I$_c$z$^{\prime}$ bands simultaneously. 
GALAPAGOS-2 was constructed to efficiently perform accurate measurements on large samples of galaxies and separate them into their components (bulge and disk) in unprecedented detail and down to fainter limits in comparison to single-band fits.
MegaMorph techniques have been successfully tested on observed ground-based and simulated data across redshifts \citep{haussler13, vika14, vika15} with an increase of the stability and accuracy of measured properties.
This is of great importance for our ground-based Subaru data and thus ideally suited for this study. 
GALFITM requires a point spread function (PSF) image, used to convolve the models to correct for seeing effects. 
This ensures that even small galaxies that are close to the PSF or pixel resolution limit can be constrained.
We used a statistical PSF generated from a median stack of bright, unsaturated stars in each frame. 

We performed both a single-component fit and a two-component S\'ersic light profile decomposition where we modeled the bulges with DeVaucoleur profiles ($n$=4) and disks as exponentials ($n$=1). 
In our setup, we ensured complete freedom for the wavelength dependence of magnitude by using a fourth-order polynomial with five coefficients, equal to the number of bands in our data set.
Following \citet{haussler13}, we constrained minimum/maximum magnitude deviation to $-5 \leq m_{fit} -m_{input} \leq 5$ where $m_{input}$ was established by calculating the offsets to the MAG\_BEST parameter of bright objects of a SEXTRACTOR run. 
Values for magnitudes were confined between $0 \leq m \leq 40$ and effective radii between $0.3 \leq r_e \leq 400$ pixels corresponding to sizes of 0.34 and 456 kpc at the redshift of this cluster and Subaru's pixel scale of 0.202 pixels/arcsecond. 
We allowed profile half-light radius and S\'ersic index to vary linearly with wavelength. 
This was chosen after examining the wavelength dependence of $r_e$ and $n$ with polynomials of higher order and finding a linear function was a sufficient and conservative option for our five band photometry.
In addition, we constrained the values for S\'ersic index $n$ to lie between 0.2 and 8 at all input wavelengths and held the centers of the bulge and disk profiles constant with wavelength.
We then selected the objects that were successfully fitted by GALFITM to obtain a clean catalog. This process excluded galaxies with parameters on or very close to a fitting constraint, which is the case for fits that do not find a minimum in $\chi ^2$ space. However, there were some objects where the fits did not move from the input parameter and did not vary in the individual bands, and some cases where the reported errors were unrealistic or very large. This lead to an exemption of 17 objects which reduced the number of galaxies from 560 to 543 for the morphological study.

In general, choosing an incorrect model may affect the size measurements. However, though intuitively we might assume that size measurements are affected by the S/N or intra-cluster light components (since the outer wings of the galaxy are not directly visible in lower S/N or crowded field images), GALAPAGOS has been shown to work well in these conditions: previous publications that addressed the robustness of structural measurements from ground based data in crowded regions, have shown that galaxies living in high-density environments do not suffer from less accurate size measurements than those in less dense regions. \citep[e.g.,][]{haussler07,lani13}. This is due to GALFIT and GALFITM's approach of minimizing the residuals through model testing in combination with sigma images. 
We will continue to address this issue and present further tests of the robustness of multi-component morphological fitting with our CLASH-HST data set (compared to simulated cluster images) in a forthcoming publication. 

For the presented study, we simply compared independent GALAPAGOS-2 measurements of effective radii, single S\'ersic index $n$, luminosities and B/T ratios from Subaru with the deeper CLASH-HST images.
This is possible because of a small overlap of the data sets in the innermost region of the cluster that allow a comparison for 80 objects with successful measurements.
Measurement uncertainties depend on brightness and size of the galaxy, with fainter and smaller galaxies deviating more strongly. This means that the typical difference of 0.12 kpc in effective radius relates to $\sim$1\% for a galaxy with 10kpc, and $\sim$30\% for a compact galaxy with a size of 0.4kpc.
The scatter in our comparison of measurements of galaxy sizes, is around 33\% or between 0.1 - 0.2 dex, depending on the apparent magnitude. Considering the large intrinsic scatter of the mass-size relation, the measurement uncertainties are therefore small.
More importantly, there is an almost one-to-one correlation - albeit with large scatter - between all of our Subaru and HST measurements, which implies that we do not introduce any systematic bias or trend towards larger or smaller sizes in our sample. 
The scatter in relation to S\'ersic index $n$ and bulge-to-total ratio somewhat larger: 53\% for S\'ersic index and up to 57\% for B/T or 0.4 dex (which amount to errors on our binned points of around 0.04 dex). However, once again, there is no systematic bias or trend towards a certain type and measurement uncertainties are independent of B/T. 
In this simple test, we find that only 10\% of the sample are classified differently according to their single S\'ersic index (15\% according to bulge-to-total ratios) with Subaru than with HST measurements. 

We performed this cross-check on a relatively small number of objects, all located in the central part of the cluster, a result of HST's small field-of-view and its centering on the Brightest Cluster Galaxy. This ultimately leads to a much higher fraction of high-$n$ galaxies in this subsample, typically more difficult to fit for any fitting routine and in a crowded area prone to host intra-cluster light. We therefore consider these number upper limits. Nevertheless, similar tests in previous work by \citet{lani13} using much larger samples of ground-based and space-based observations present comparable correlations and agree well with our measurements.
A number of publications have assessed the robustness of galaxy profile fitting procedures of the used software using large datasets of simulated data. These tests have shown that structural parameters using high quality ground based imaging can be determined accurately \citep[e.g.,][]{haussler13, guo09}, even for for distant galaxies  \citep{krywult17}.
In addition, the level of the scatter of the mass-size relation presented in this paper is comparable with the scatter of previous publications \citep[e.g.,][]{shen03, vanderwel14, cebrian14} which further indicates that the scatter of our measurement uncertainties does not introduce an extra spurious effect.
\begin{figure}
    \centering
    \subfloat{{\includegraphics[width=\columnwidth]{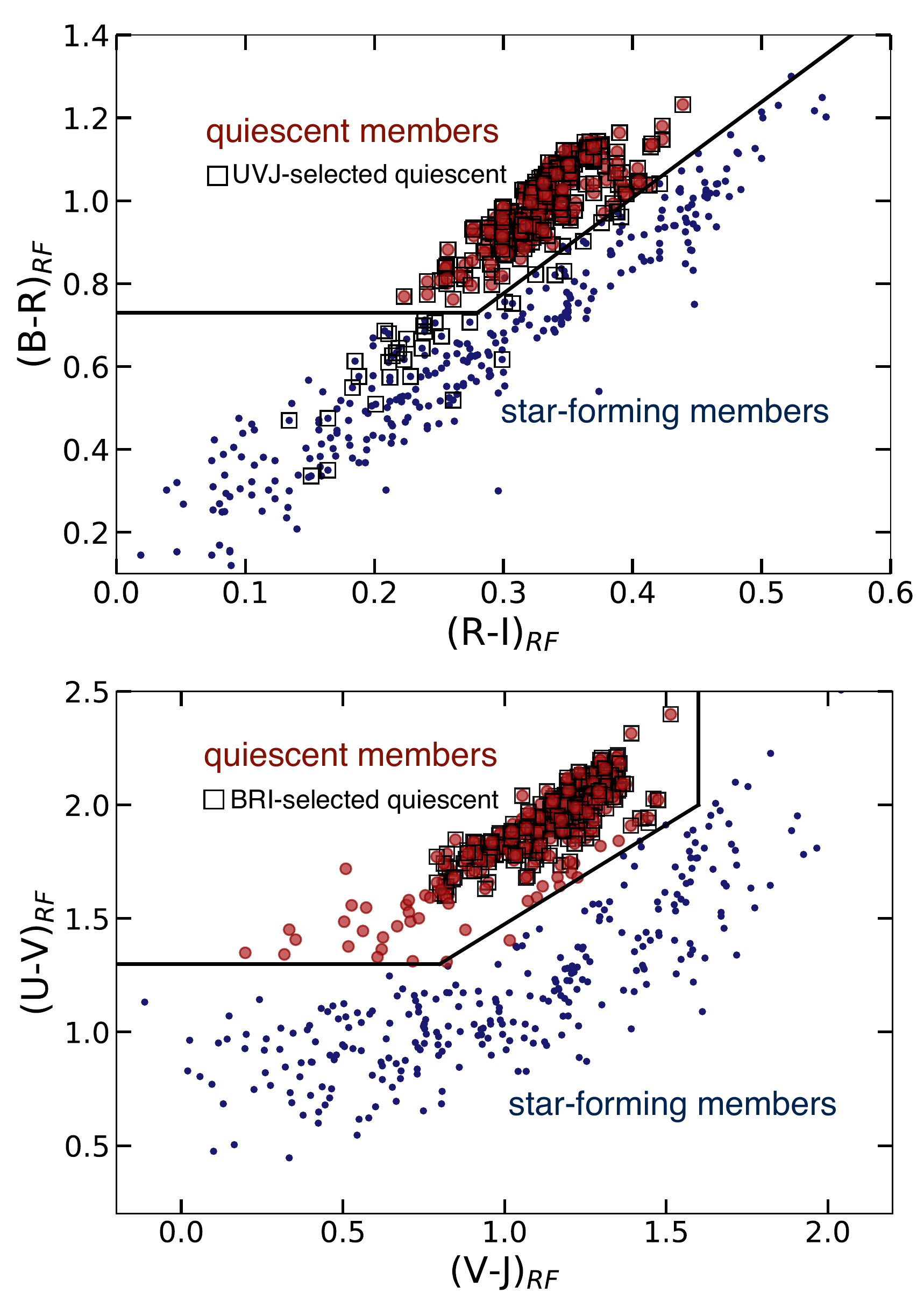} }} 
    \caption{Color-color selections of cluster members into quiescent (red circles) and star-forming (blue dots) galaxies. Top: selection according to restframe (B-R) vs (R-I) colors. Galaxies that are classified as quiescent in the UVJ diagram are marked with black frames. Bottom: Selection using restframe (U-V) vs (V-J) colors. Because our data lack near infrared observations, J band has been extrapolated synthetically for this plot. Galaxies highlighted with black frames are classified as quiescent in the BRI diagram. We do this to demonstrate that our BRI selection compares reasonably well with the UVJ diagram used in \citet{vanderwel14}.
    }
    \label{fig:color_color}%
\end{figure}

\subsection{Determination of stellar masses}
\label{sec:masses}

Stellar masses were estimated from point-spread function matched Subaru photometry with a  photometric accuracy of $\sim 0.01\,$mag in all five passbands (see also \citet{umetsu12}). 
In order to derive consistent colors, photometric catalogs were created with SEXTRACTOR \citep{bertin96}, using PSFs constructed from a combination of 100 stars per band. 
To calculate stellar masses and restframe absolute magnitudes, we used the code \textit{LePhare} \citep{arnouts11}. This fits SEDs based on stellar population synthesis models from \citet{bruzual03} to our multi-wavelength photometry, assuming a Chabrier IMF \citep{chabrier03}. 
We fit templates with an exponentially declining star formation history with $\tau = 0.1 - 30\ \mathrm{Gyrs}$ and  attenuated by a \citet{calzetti00} dust law with extinction E(BV) = 0--0.5 magnitudes. 
Uncertainties correspond to the 68\% confidence interval; marginalized over extinction, star-formation rates, age and metallicity. 
At this redshift, (B-R$_c$) brackets the 4000\AA\ break, sensitive to galaxy mass-to-light ratios, and therefore ensures robustness of our estimated masses.
To test for possible biases, we compare stellar masses from an independent sample of galaxies with 5-band photometry to measurements from 11 bands, suppress bands, and change extinction laws and internal parameters. 
Overall we find that stellar mass is a robust parameter. The average change is minimal within the scatter, around 0.1 dex, generally smaller than the typical 68\% confidence intervals from \textit{LePhare}.

\subsection{Classifying the cluster galaxies}
\label{sec:classification}

The mass-size -- relation depends strongly on galaxy morphological type (disk-dominated/spheroid-dominated galaxies) and star-formation rate (star-forming/quiescent galaxies). 
It is therefore sensible to consider the size distribution for samples of different types separately. This is especially true for galaxies in clusters, as strong variations in type \citep[e.g.,][]{dressler80, kauffmann03, treu03, brinchmann04, wilman09,vulcani15} and star-formation fractions \citep[e.g.,][]{balogh04, pasquali09, peng10, nantais13, muzzin14} with environment have been found. In dense environments, the usual correspondence between structure and star-formation breaks down. Studies find that disk galaxies show a stronger dependence on environment and become redder for higher densities \citep{bamford09, wolf09, lopes16}.  
It is therefore important to consider galaxies both classified according to their morphological type as well as their star-formation activity. 

To motivate our study, we first wish to compare our cluster galaxy measurements with field measurements from \citet{vanderwel14} using data from the CANDELS survey. 
Following their approach, we therefore match the classification of our galaxies into star-forming and quiescent galaxies depending on their location in color-color diagrams \citep[e.g.,,][]{wuyts07}. 
Briefly, in these diagrams galaxies populate two different sequences, an old-age sequence of quiescent galaxies and a star-forming sequence of galaxies with stronger star-formation rates and higher dust contents \citep[e.g.,,][]{whitaker13}. 
These regions are chosen arbitrarily by previous authors with the main criteria that the boxed galaxies enclose the easily distinguished passive population. 

In their paper, \citet{vanderwel14} use the popular UVJ-diagram to separate galaxies into star-forming and quiescent populations.
However, our optical 5-band Subaru photometry does not include near infrared data. We therefore extended our wavelength range synthetically by producing J-band magnitudes from Bruzual \& Charlot (2003) models with \textit{LePhare} to compare the star-forming/quiescent selection with the less commonly used BRI-selection plots.
We also verify the consistency of quiescent galaxies in the BRI-color plot with stellar population model tracks of a passively evolving galaxy with formation time at z=4.
While the location of quiescent galaxies in both selection diagrams are roughly consistent, the black open squares in Fig. \ref{fig:color_color} demonstrates where the galaxies classified as quiescent lie in the respective other identification plot. 
We notice (and investigate) a difference among the classification diagrams: more galaxies are classified as quiescent in the UVJ- than in the BRI-diagram. They are likely a mix of passive and actively star-forming galaxies.
An inspection of the morphologies of these galaxies reveal that these are often disky galaxies.
The UVJ-diagram was designed to separate pure quiescent galaxies from galaxies with larger amounts of dust (i.e., red star-forming galaxies), typically less massive, which explains this discrepancy. 
This difference, however, only plays a minor role in the general trends of the mass-size diagrams. Because of their mixed assignments to types, these galaxies do not have a preferred location in the MSR; they evenly spread in-between the star-forming and the quiescent relations. 
We find no change in the overall results of the MSR when we interchange color-selections and therefore show results for galaxies classified as star-forming/quiescent as defined by their BRI-colors. 
   \begin{figure} 
      \centering
        \includegraphics[width=\columnwidth]{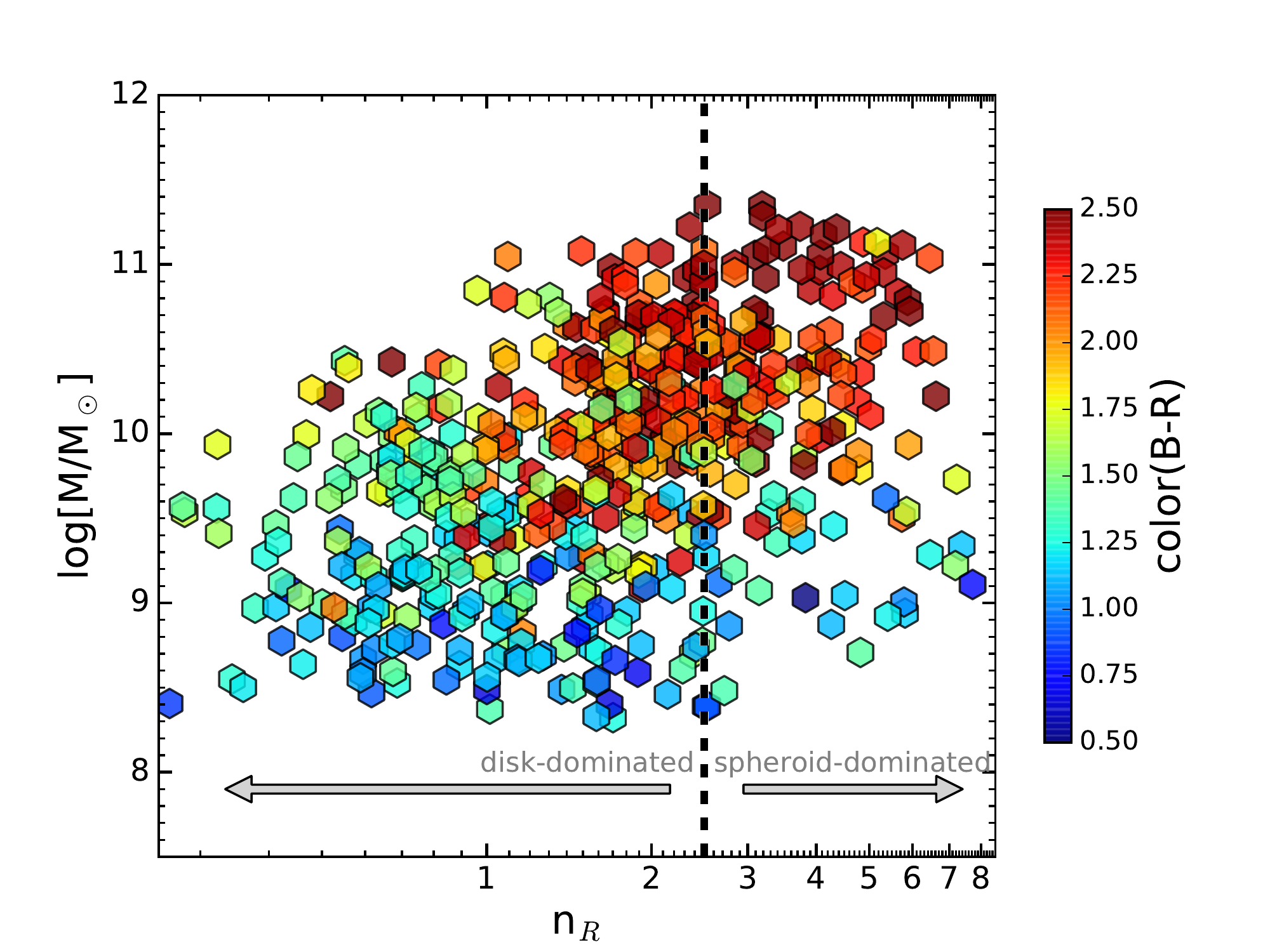}
   \caption{The 3D distribution of the sample: S\'ersic index $n$ versus stellar mass, color coded by their (B-R) colors. The dashed line indicates the hard cut in S\'ersic index ($n$=2.5) used here for the first morphological classification into disk-dominated and spheroid-dominated galaxies. }
        \label{fig:n_mass_color}
    \end{figure}
\bigbreak
We also study the MSR of our sample as a function of galaxy type as classified by their structure. We first divide the galaxies by their single profile S\'ersic index $n$ which describes the shape of the radial light profile, or concentration. 
The recent literature has established $n<2.5$ and $n>2.5$ as the common separator to distinguish between disk-dominated and spheroid-dominated galaxies respectively \citep[e.g.,][]{shen03,barden05, mei12, huert-comp13,graham13,cebrian14, lange15}.
This division has been found to roughly correlate with quiescence \citep[e.g.,,][]{devaucouleurs48} and the dominance of a spheroidal component \citep{bruce14}. 
However, $n$ depends on mass for early-type galaxies, so care is required. 

Stellar mass maps and colors do not project uniquely into morphology. All distributions of structural and stellar population parameters used for morphological determinations are washed out by a considerable scatter.
In our case, the cluster environment increases this scatter. For example, from observations in nearby clusters, we know that even grand spiral design galaxies are often red and passive \citep{wolf09}. 
A non-negligible fraction of $n$<2.5, disk-dominated cluster galaxies are red and some $n$>2.5, spheroid-dominated galaxies are blue, as seen in Fig. \ref{fig:n_mass_color}.
In our sample of MACS1206 galaxies, 22\% of the red quiescent population of our cluster galaxy sample have $n$<2.5, i.e., are dominated by a disk. 

\bigbreak
Almost all existing studies of the environmental dependence of galaxy sizes have been executed using single S\'ersic light profile fits to measure structures and effective radii.
However, it is evident that galaxies falling into a cluster environment undergo dramatic structural transformations, changing from star-forming disk-galaxies to passive bulge-dominated systems. 
For a comprehensive study of galaxy evolution in clusters, we therefore gain essential knowledge from tracing the individual galaxy components, bulge and disk, by fitting two-component models.
At the simplest level, this approach allows us to classify galaxies according to their bulge-dominance. 
The systematic behavior of B/T ratio reflects the main hierarchical galaxy formation mechanisms \citep[e.g.,][]{stein_navarro02} and therefore defines the position in the Hubble tuning fork diagram.

To calculate the bulge-to-total ratio, we consistently use the two-component fits for every galaxy in our sample. 
We use the band with the highest S/N, the $I_{c}$ band, also chosen as the primary band on which all deblending and masking decisions are made in GALAPAGOS-2. 
We note that the number of galaxies in different bins only changes slightly (5\% at the most) if we change this to measurements of the $V$-band.

Today we know that the great majority of galaxies are made up of multiple components. 
However, because of imperfect data quality, there are cases where a single S\'ersic fit is the better representation of the light.
While we acknowledge that there may be statistical ways to determine whether a galaxy is best represented by 1- or  2-component models, this also introduces an artificial dichotomy into the population. The approach in this paper of fitting every galaxy with 2 components assures consistency by treating all galaxies in our sample in a homogeneous way.
For some galaxies, the fitting of two components failed. Reasons could be neighboring sources that were not masked properly or because the galaxy simply does not have two components. In these cases, the bulge measurements exceed any reasonable number and these objects (12\% of our sample) are therefore excluded from the discussion and any figures that use B/T measurements. 
   \begin{figure} 
      \centering
        \includegraphics[width=\columnwidth]{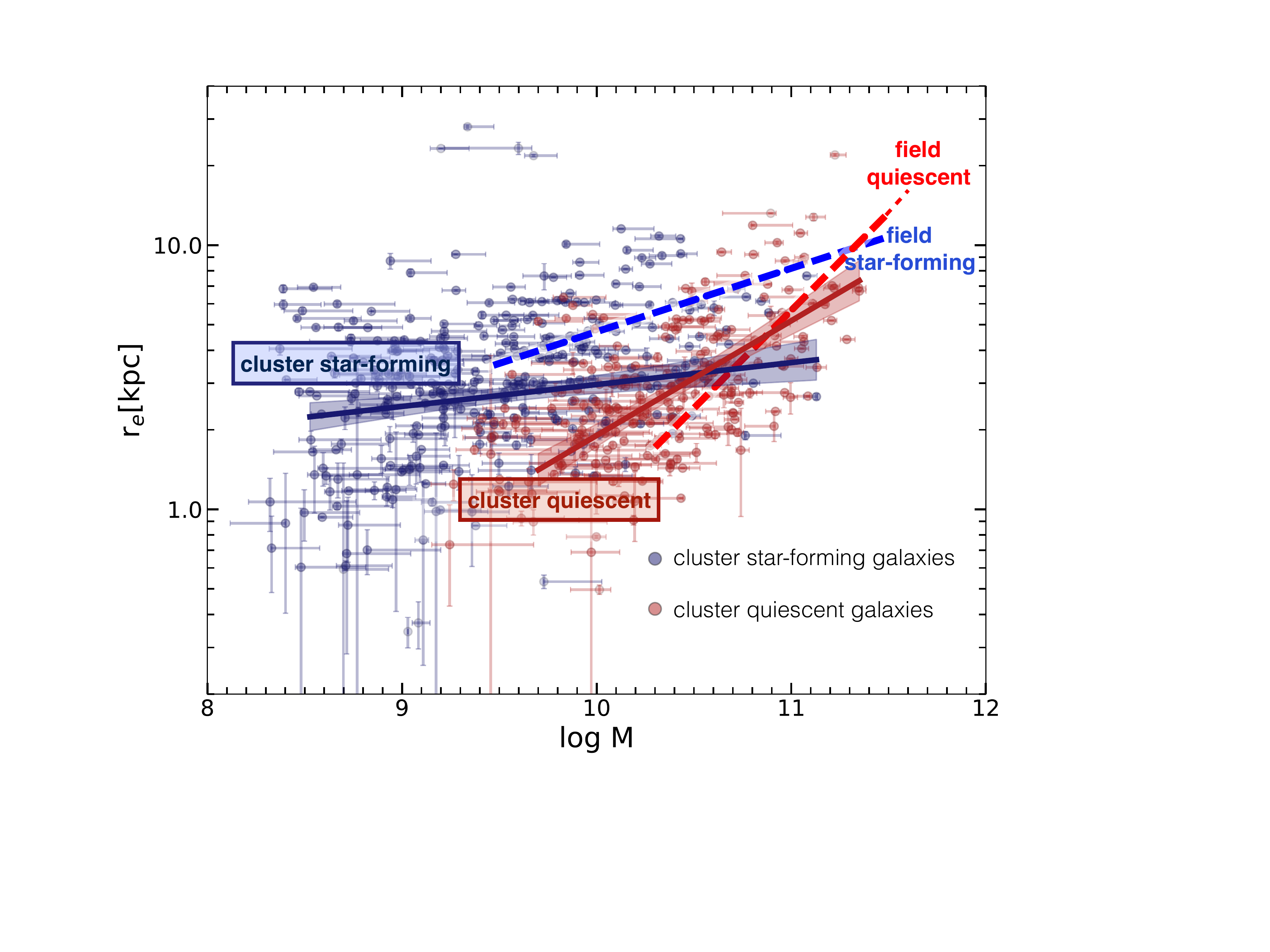} 
   \caption{Comparison of stellar-mass -- size relations of cluster galaxies in MACS1206 (solid lines) to field galaxies from \citep{vanderwel14} (dashed lines). This simple comparison suggests some minor differences.
   Points are BRI-selected star-forming (blue) and quiescent (red) cluster galaxies. The blue and red solid lines are single power-law fits to the entire respective weighted data $\log (\mathrm{M/M}_{\odot}) >$ 8.5 for star-forming and  $\log (\mathrm{M/M}_{\odot}) >$ 9.2 for quiescent galaxies 
   (see Appendix \ref{sec:appendix}).
   Error bars on the points indicate the 68\% confidence level on the mass from SED fitting as well as (statistical) errors on the effective radii provided by GALFITM. Including these errors in the fitting routine does not change the result significantly and we indicate this change with the shaded area.}
        \label{fig:MSR_BRI}%
    \end{figure}
\section{Results: The size distribution of cluster galaxies}

Galaxy sizes obey a log-normal distribution $\mathrm{N}(\log\,r, \sigma_{\log\,r})$, 
where $r$, and potentially $\sigma$, are a function of galaxy mass.
To quantify the observed MSR, we first fit them with analytic formulae, motivated by \citet{vanderwel14} and \citet{shen03}'s earlier work. Both authors found a single power law of the form
\begin{equation}
\log R_e = a + b \log \frac{M}{M_{\odot}}
\end{equation}
to be a good fit describing the data and its usage was recommended for canonical references \citep{lange15}.

However, calculating the average sizes in bins of mass offer a complementary description of the MSR that allows us to analyze more subtle differences that may be attributed to cluster mechanisms, the main concern of this investigation. 
To quantify the size distribution of galaxies, we therefore show both approaches for our analysis. 

(i) Power laws are shown in Fig. \ref{fig:MSR_BRI} as solid lines for star-forming and quiescent galaxies of the cleaned data. 
The linear least square fits are performed on data where obvious outliers to the fit were removed (i.e., 5 galaxies lie significantly above the general relation) and errors in stellar masses and effective radii considered.  

Our data allow mass limits down to  
$\log (\mathrm{M/M}_\sun$) = 8.5 for late types and $\log (\mathrm{M/M}_\sun$) = 9.2 for early type galaxies.
Observational limits do not permit conclusions of the very faint quiescent galaxy population. For MACS1206, the completeness for early type galaxies starts to drop below $10^{9.2} M_{\odot}$. (We refer to Sec. \ref{sec:measurements} and the Appendix for our discussion and figures on the sample and incompletenesses).
We show fit lines to field data (described in more detail in section \ref{sec:MSR_field_cluster}) as dashed lines in Fig. \ref{fig:MSR_BRI}. 

(ii) The second - and principal - approach to quantifying the size distribution in MACS1206 uses weighted averages. In all subsequent figures that feature the MSR, we show the mean sizes in 0.5dex mass bins. The bins are connected by lines, and shaded regions indicate the $1\sigma$ error on the weighted averages.
In this way, we ensure a direct comparisons with field data from \citet{vanderwel14} that use the same 0.5dex mass bins. We show their field results as triangles in Fig.\ref{fig:MSR_vdwbins}. 
\citet{vanderwel14} used a slightly different cosmology from us, so the reported sizes in our paper are $\sim$3\% lager than in their work, a difference that it completely buried in the relatively large scatter of the MSR and also smaller than our typical error bars.

\subsection{Comparison of galaxy sizes in the cluster to the field}
\label{sec:MSR_field_cluster}

\begin{figure} 
      \centering
        \includegraphics[width=\columnwidth]{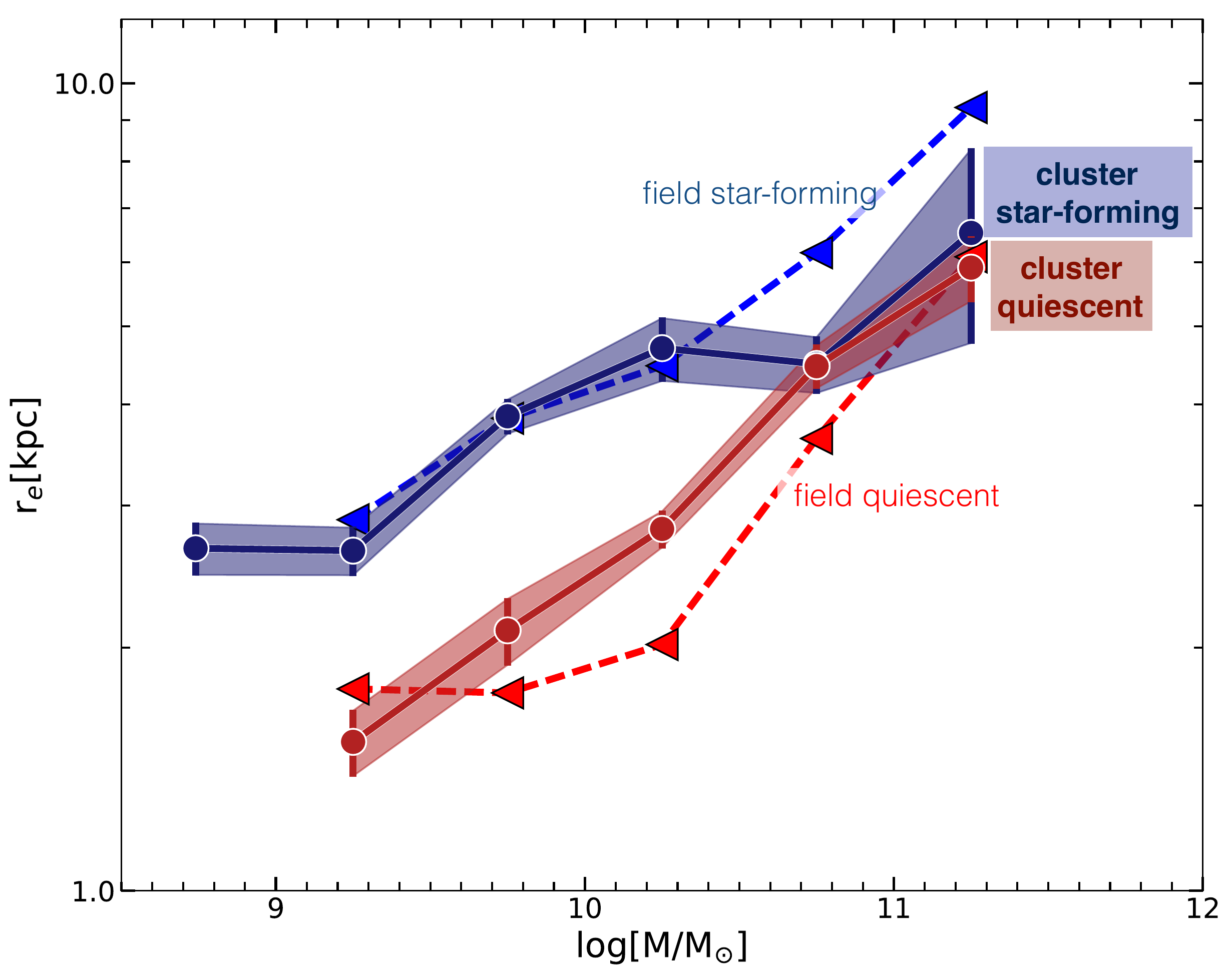}
   \caption{Comparison of the stellar-mass -- size relations in field and MACS1206 cluster environments for star-forming (blue) and quiescent (red) galaxies and their errorbars. We divide the sample into mass bins for a more detailed description of the sizes. The solid lines show the weighted mean sizes in 0.5dex mass bins of both samples. The shaded regions indicate the 1$\sigma$ error on the mean. The red and blue triangles are median points for field galaxies from \citet{vanderwel14}, their Fig. 8, interpolated to our redshift and connected by dashed lines.}
        \label{fig:MSR_vdwbins}%
    \end{figure}

Before concentrating on the cluster galaxies themselves, we first compare the sizes of cluster members with those of field galaxies provided in \citet{vanderwel14} who measured S\'ersic half-light radii with GALFIT for a large sample of galaxies in the CANDELS field.
They present the stellar-mass--size relation both as redshift-dependent linear functions and binned medians at six different redshift bins.
In Figs. \ref{fig:MSR_BRI} and \ref{fig:MSR_vdwbins}, these are interpolated to our redshift and represented by red (quiescent) and blue (star-forming) dashed lines and triangles.

Galaxies possess significant radial color gradients. Their sizes are therefore known to depend on wavelength \citep[e.g.,][]{kelvin12, vulcani14}. For this reason, \citet{vanderwel14} present their MSRs corrected to a restframe wavelength of 5000\AA. This in turn demands the same choice of waveband for our measurements of effective radii, i.e., the V band (the band including the wavelength of 5000\AA) in order to minimize the chance of misinterpreting any possible differences.
Contrary to \citet{vanderwel14}, we do not take a mis-classification of 10\% into account. Instead, we opt for showing the mass-size relation for different classifications to directly highlight any discrepancies resulting from various approaches reported in the recent literature.

We show the stellar-mass -- size relation for quiescent (in red) and star-forming (blue symbols) MACS1206 cluster galaxies as defined by their BRI color space (see Section \ref{sec:classification}) in Fig. \ref{fig:MSR_BRI}.
We plot a power law to all galaxies above our mass limits of $\log(\mathrm{M/M}_{\odot}$)=8.5 for late types and $\log(\mathrm{M/M}_{\odot}$)=9.2 for early type galaxies.

At all mass bins, quiescent galaxies are smaller than star-forming galaxies for a given mass, except at highest stellar masses covered by our sample (galaxies of $\sim 10^{11}M_{\odot}$), where we see that the quiescent and star-forming populations converge. At very high stellar masses ($>2 \times 10^{11}M_{\odot}$), the two populations have similar sizes or reverse their size behavior.  
At first glance, this simple approach suggests a striking trend for cluster star-forming galaxies to be smaller than their field counterparts at same mass. 
Sizes of quiescent galaxies appear to be comparable or larger in the cluster than in the field environment. 

This popular procedure of comparing linear fits obviously over-simplifies the task at hand and can only hint at differences at the high mass end of galaxies. 
Clearly, a more detailed analysis is needed to adequately investigate any indications for a trend of decreasing sizes of star-forming galaxies in clusters. 

\bigbreak
In Fig. \ref{fig:MSR_vdwbins} we therefore show the direct comparison of galaxy sizes from our cluster data (solid points and lines) to the CANDELS field measurements (triangles and dashed lines) in the same bins of mass as specified by \citet{vanderwel14}.
Errors on the mean (which we refer to as $\sigma$ in the following) for the much larger field study are naturally smaller than for our cluster sample. Instead, \citet{vanderwel14}, show the standard deviation as 16th and 84th percentiles in their plots. Within the 16th and 84th percentiles of either data set, the field and cluster relations (i.e., the clouds of points) naturally overlap, due to the large scatter inherent to the mass-size relations for star-forming and quiescent galaxies.

Comparing their underlying trends in Fig. \ref{fig:MSR_vdwbins}, we see the same overall behavior for galaxy sizes in the cluster as in the field, i.e., the trends of cluster galaxies generally follow the field relations for quiescent and star-forming galaxies.
Looking more closely, Fig. \ref{fig:MSR_vdwbins} reveals some discrepancies for average sizes of intermediate quiescent and high mass star-forming galaxies.
First, the relation for star-forming galaxies with masses greater than $10^{10}\ \mathrm{M}_{\odot}$ deviates from the field relation towards smaller sizes. When comparing at fixed stellar mass and in the same bin sizes as \citet{vanderwel14}, the average sizes of star-forming cluster galaxies for the two highest mass bins (i.e., between 10.5 and 11.5 log[M/M$_{\odot}$]) are more than 3 and 1 $\sigma$ below the respective bins reported in the field.

The mass-size -- relations of quiescent galaxies show the same general behavior independent of their environment, i.e., quiescent galaxies are smaller than star-forming galaxies.
The residuals of cluster bins to the median field, however, reveal that quiescent galaxies of the intermediate mass bins are between 2 and 5 $\sigma$ larger in the cluster than in the field.

\bigbreak
Here, we conclude the direct comparison of the size distribution for cluster and field galaxies. We saw that their mass-size -- relations for star-forming and quiescent galaxies show similar overall trends over wide mass-ranges.
While their clouds of points overlap, an obvious offset of the mean relations in some places is apparent. We find indications for smaller sizes of high mass star-forming galaxies in clusters, and for larger quiescent galaxies at intermediate masses. We did this by contrasting the mean sizes of cluster galaxies (and their errors) to those of field data at same stellar mass bins. The preceding section compared measurements in different data sets that were analyzed by different authors, so naturally, some care is  required in interpreting these results. However, note that the same software has been used, with comparable setups, and with a careful match of wavelength, redshift, and mass-bins. We therefore believe that our comparison is fair.
For the rest of the paper we concentrate on the cluster members within our own data set and work towards uncovering possible reasons for these variations.

\subsection{The stellar-mass -- size relation of galaxies in MACS1206}
\label{sec:insidecluster}

While the key observables galaxy color and structure are well correlated in the field, this dichotomy breaks down for dense environments. 
The cluster environment boosts galaxy transformation from blue to red colors and from disk-dominated to spheroid-domianted galaxies with a higher fraction of transition objects.
For example, we know that in clusters quenched, red galaxies are not uniquely associated with high-S\'ersic index $n$ galaxies. We therefore want to investigate if there are any differences in the MSR that arise as a result of this transformation process inside the cluster.
   \begin{figure} 
      \centering
        \includegraphics[width=\columnwidth]{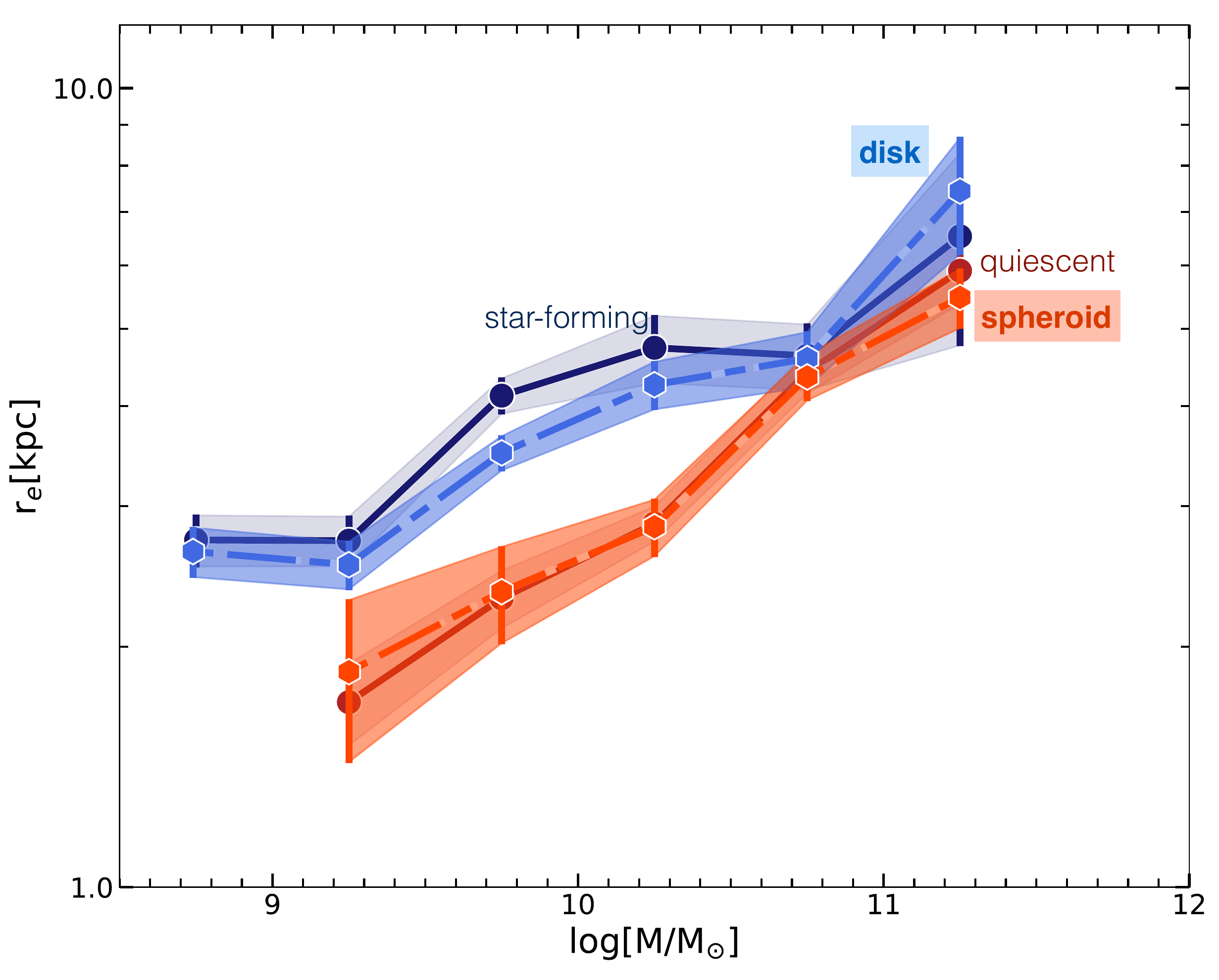}
   \caption{Comparison of sizes for cluster galaxies divided by their structure (dot-dashed lines) and by their color (solid lines). Mean sizes agree well. Only at intermediate masses, disk-dominated galaxies are slightly (1-2 $\sigma$) below size measurements of star-forming galaxies (in blue colors).
   Lines show the weighted mean sizes in 0.5dex mass bins of the clipped cluster data and shaded regions indicate the 1$\sigma$ error on the mean.
   }
        \label{fig:MSR_morph_vs_sf}%
    \end{figure}
   \begin{figure} 
      \centering
        \includegraphics[width=\columnwidth]{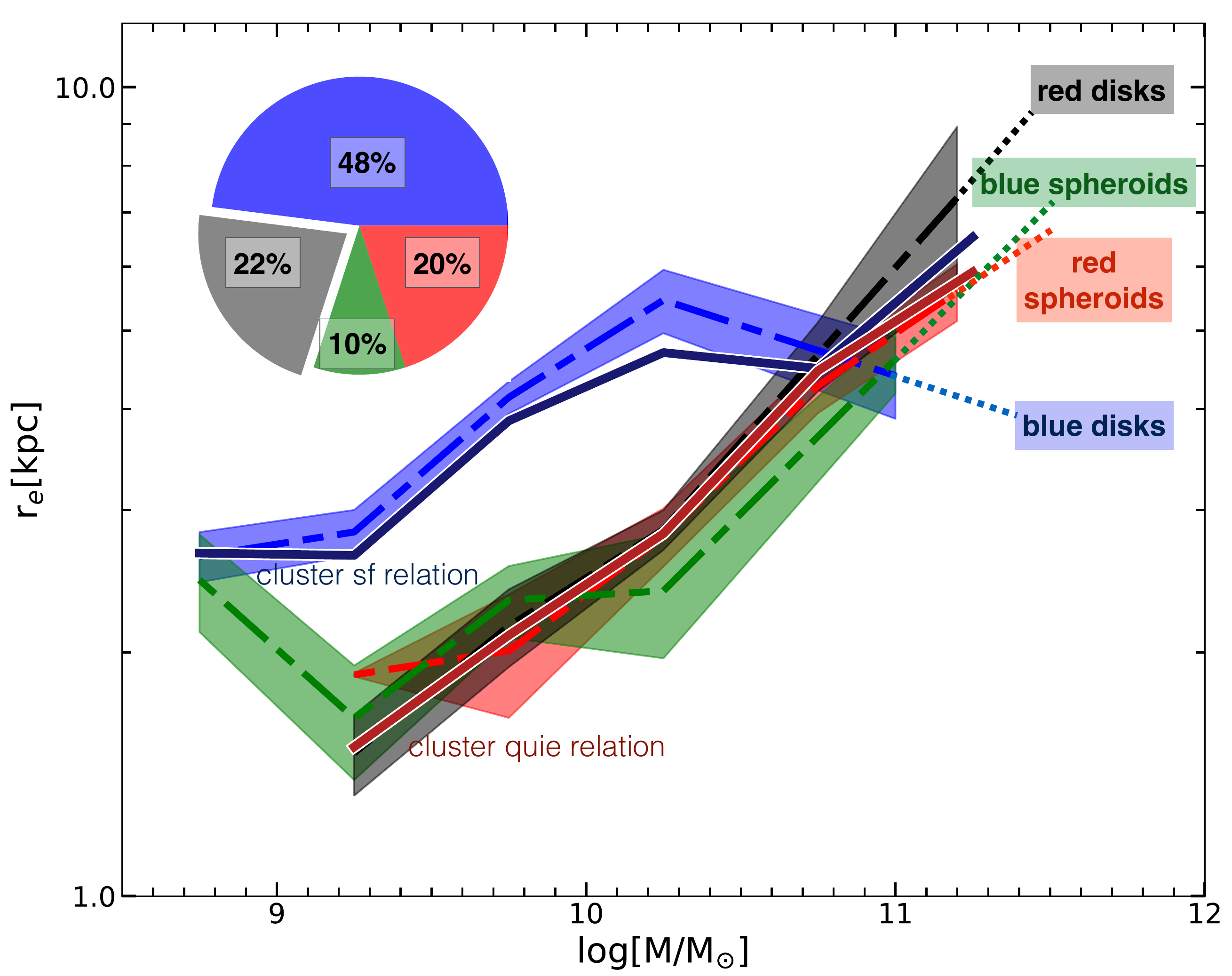}
   \caption{The stellar-mass -- size relation for star-forming galaxies with $n<2.5$ (blue), star-forming galaxies with $n>2.5$ (green), and quiescent galaxies with $n>2.5$ (red), quiescent galaxies with $n<2.5$ (black). Star-forming disk galaxies are much larger than all other sub-classes, just above the general cluster relation for star-forming galaxies. Solid lines show the weighted mean sizes in 0.5dex mass bins of the clipped data. The shaded regions indicate the 1$\sigma$ error on the mean. We also show the cluster relation for star-forming and quiescent galaxies presented in Fig. \ref{fig:MSR_vdwbins} as solid lines for comparison and the distribution of the types in percentages.}
        \label{fig:MSR_combined}%
    \end{figure}
\subsubsection{The mass-size relation in clusters is a composition of diverse populations}
\label{sec:comp}

The following section compares the mass-size--relations of star-forming and quiescent cluster galaxies and $n<2.5$ disk-dominated and $n>2.5$ spheroid-dominated cluster galaxies.
In Fig. \ref{fig:MSR_morph_vs_sf} we show the size distribution for disk- and spheroid cluster galaxies as defined by their single S\'ersic fit.
As in the previous plot (Fig. \ref{fig:MSR_vdwbins}), the samples are further divided into 0.5dex mass bins for which the weighted averages were computed on the clipped sample. 
The general trend of disk-dominated galaxies is similar to the star-forming one, the one of spheroid-dominated galaxies to that of quiescent galaxies. (Dot-dashed lines represent galaxies divided by S\'ersic index, solid lines repeat the trends of galaxies classified according to their star-formation status as seen in Fig. \ref{fig:MSR_vdwbins}.) 

The majority of star-forming galaxies are also galaxies with low S\'ersic index, so a very similar trend is not surprising. The small discrepancy between the general trends of disk galaxies and of star-forming galaxies $1$ to $2\sigma$ at most) is likely due to a number of quiescent disk-dominated galaxies (e.g., passive ``red disks'').
This type of galaxy is classified quiescent in one selection and disk-dominated in the other. 
More specifically, 361 galaxies have $n$<2.5 (i.e., are classified as disk-dominated galaxies), whereas only 313 are star-forming galaxies. For these ``red disks'', the correspondence between galaxy morphology and color as we know it is no longer valid.
This population of galaxies requires a physical process that shuts down star-formation but leaves the disk in place - at least for a significant period of time. 
The existence of a ``transitioning population'' of galaxies where color and morphology do not match our expectations, suggests different timescales for shutting down star-formation and turning disks into spheroids. 

The trends for quiescent and high S\'ersic index galaxies agree within the errors at all masses. 
   \begin{figure} 
      \centering
        \includegraphics[width=\columnwidth]{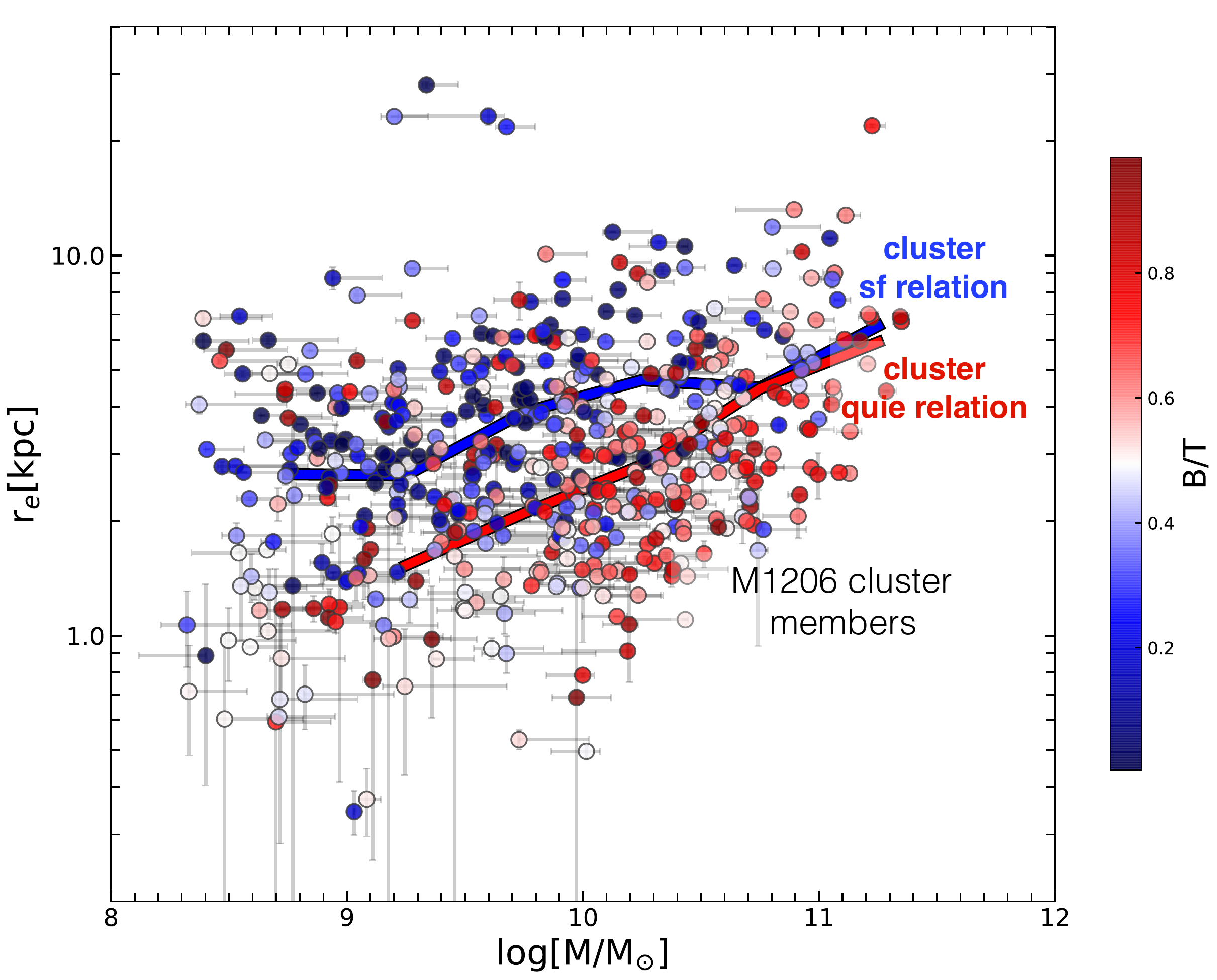}
   \caption{We show the dependence of the size distribution of cluster galaxies on B/T. The plot highlights that the structure of the galaxy, i.e., the strength of the bulge, influences the location of the galaxy in the mass-size plane. Galaxies with lowest B/T have greater sizes, clustering around and above the relation for star-forming cluster galaxies (blue line), galaxies with highest B/T lie on and below the quiescent cluster relation (red line).} 
        \label{fig:MSR_BT}
    \end{figure}
   \begin{figure} 
      \centering
        \includegraphics[width=\columnwidth]{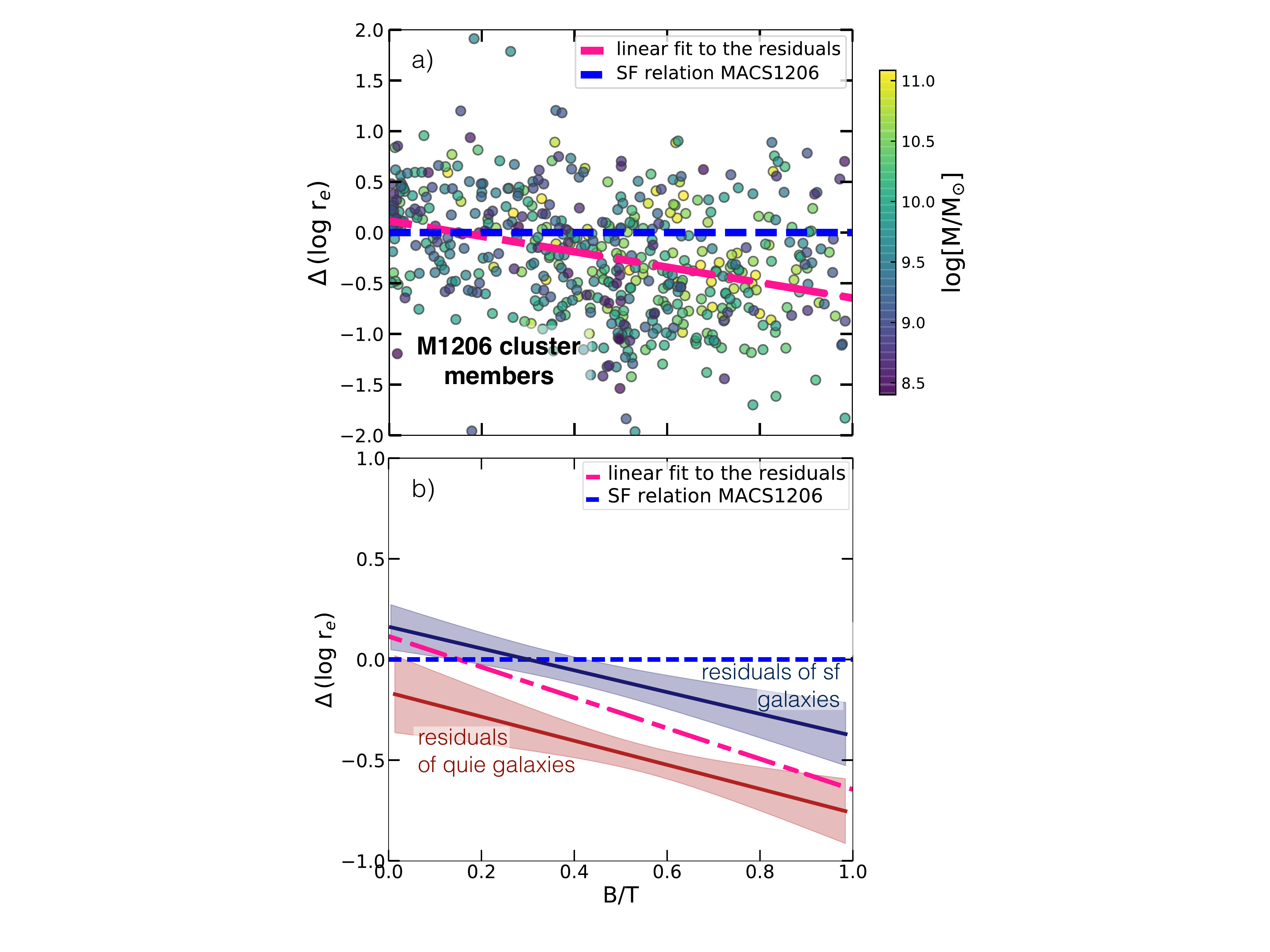}
   \caption{Residuals from the star-forming relation vs. B/T. The plots quantify the change of bulge-to-total ratio for cluster galaxies. For each galaxy we calculated the difference in size from the relation defined by the star-forming cluster galaxies (represented by blue dashed lines) and plotted these residuals as a function of B/T. 
   a) The top panel shows all galaxies, color-coded by their mass, to preserve this information. The pink dot-dashed line is the linear fit to all galaxies. Galaxies with higher B/T deviate most from the SF line and thus have smaller sizes. 
   b) The bottom panel shows the linear fits (and their error bands) to star-forming (blue) and quiescent (red) galaxies. We see that at same B/T, SF galaxies are larger than QUIE galaxies. Furthermore, at higher B/T, the average behavior oscillates from the SF to the QUIE fit demonstrating that at low B/T, the larger sizes are dominated by star-forming galaxies whereas at high B/T, smaller sizes are attributed to quiescent galaxies. }
        \label{fig:MSR_hubble}%
    \end{figure}
\bigbreak
To further investigate the cluster population of objects transitioning from blue star-forming to red quiescent galaxies, we break down the MSR in Fig. \ref{fig:MSR_combined} into star-forming galaxies with low (blue line), and high S\'ersic index (green line), and quiescent galaxies with low (black line), and high S\'ersic index (red line). 

The (blue) star-forming $n<2.5$ population (in total 48\% of our sample) lies just above the relation of all star-forming galaxies in the cluster; all other types have significantly smaller sizes.  
Quiescent spheroid-dominated galaxies (red) make up 20\% of the cluster sample; their sizes are consistent with the overall quiescent population.
In addition to these ``standard'' types of the galaxy bimodality, we find systems that do not identify with either category. These are galaxies that are blue in color, i.e., star-forming, but spheroid-dominated in structure; or they are red in color, i.e., quiescent, but disk-dominated.
They might present a link between quenching of star-formation and morphological evolution of galaxies. Different to the MSR of field galaxies, the MSR for cluster galaxies will therefore always include a greater mixture of galaxies at diverse transitional stages.

The green line represents a small population (10\% in our sample) of of spheroid-dominated galaxies that are still forming stars. 
These galaxies are understood to be a mixture of recent newcomers to the red sequence \citep{ruhland09}, or rejuvenated spheroid-dominated galaxies that have experienced a recent period of star-formation after minor-mergers \citep{kaviraj09} or in gas-rich environments \citep{kannappan09}. 
In our sample, this type of galaxy is considerably smaller than their blue low-$n$ counterparts, often as small as quiescent galaxies.

Fig. \ref{fig:MSR_combined} reveals a surprisingly abundant population of quiescent galaxies with low S\'ersic index: 22\% of our sample are ``red disk'' galaxies (black line). 
At intermediate to lower masses, where this type of galaxy is most abundant, they have small half-light radii, comparable to ``traditional'' quiescent spheroid-dominated galaxies. 
This means that, even though they are disk galaxies, their size distribution resembles that of spheroid-dominated galaxies. 

This observation leads to the fundamental question: Why are the global size measurements of quiescent galaxies smaller than for star-forming galaxies, seemingly independent of their structure as defined by their single S\'ersic index? 
Or asked in a different way: What makes star-forming disk-dominated galaxies become smaller when they become quiescent? 
We investigate this question by analyzing the bulge and disk components of our cluster galaxy sample.
First, we show the galaxy population with increasing bulge-to-total ratio (see Section \ref{sec:measurements} for the definition of our classification) to test if an increase of bulge dominance can be held responsible.
We then plot the sizes of only the disk-component for star-forming and quiescent galaxies separately to examine whether the disk is becoming smaller in quiescent galaxies.

\subsubsection{Galaxy sizes depend on internal structures and star-formation activity} 
\label{sec: MSR_hubble_all}

We now turn to our two component measurements where we separate the galaxies into light contributions from bulges and disks to distinguish between galaxies with increasing B/T-ratios. 
This approach will give us valuable insight into processes that involve the bulge.

Fig. \ref{fig:MSR_BT} shows the mass-size relation for all cluster galaxies
color-coded by their bulge-to-total ratio. It becomes evident that the mass-size relation systematically depends on the strength of the bulge. 
It shows galaxies smoothly decreasing in size as the bulge becomes more dominant. 
Rather than showing a clear binary opposition (star-forming/quiescent, disk-dominated/spheroid dominated), we see how the size distribution is gradually changing with increasing B/T. 
Systems close to the cluster relation of star-forming galaxies are represented by low B/T, i.e., late type disk galaxies, whereas high B/T (early type) spheroid-dominated galaxies populate the region around the steeper quiescent relation. 

The sizes decrease naturally as the dominance of the more compact spheroidal component increases.

The top panel in Fig. \ref{fig:MSR_hubble} expresses this information quantitatively: for each object we plot the difference of their effective radius to the relation established by the star-forming cluster galaxies, versus B/T. The pink dot-dashed line presents their linear fit. Galaxies with larger B/T values, and therefore greater dominance of the bulge, are farther away from the star-formation relation, which means, they are smaller. The color-coding shows that, in general, they are also higher in mass.

In the lower panel of Fig. \ref{fig:MSR_hubble}, we split the sample into star-forming and quiescent galaxies, again plotting their residual size to the star-forming (SF) cluster relation against B/T. 
While we see the same trend in the x-direction, that is, of galaxies with increasing B/T (low B/T galaxies have larger sizes, higher B/T galaxies increasingly smaller sizes), there is a striking difference between star-forming and quiescent galaxies: 
independent of their B/T ratio and at a given mass, star-forming galaxies have larger sizes than quiescent galaxies. 
More specifically, star-forming galaxies \textit{of all B/T} are generally close in size to the overall cluster star-forming population; quiescent galaxies, independent of their B/T, are smaller, and thus closer to the relation established for quiescent cluster galaxies.

In quiescent galaxies, the bulge is more dominant than in star-forming galaxies, and a prominent bulge naturally leads to smaller sizes. 
In addition, the figure illustrates that if we concentrate on star-forming and quiescent galaxies separately, there is still a conspicuous difference in size  -- at same B/T.
The figure thus conveys that B/T alone cannot explain the picture presented so far.
Bulge dominance, therefore, can only be partly responsible, and other mechanisms connected to star-formation activity must contribute as well.

Our cluster sample evidently reveals a difference in sizes between star-forming and quiescent galaxies at fixed B/T and mass. The most readily available explanation for this is that the disk component of quiescent galaxies was affected by the cluster environment. In the next section we will therefore examine whether quiescent galaxies have smaller disk components, perhaps due to an outer fading or truncation of the star-forming disks inside the cluster.

    \begin{figure} 
    \centering
        \includegraphics[width=\columnwidth]{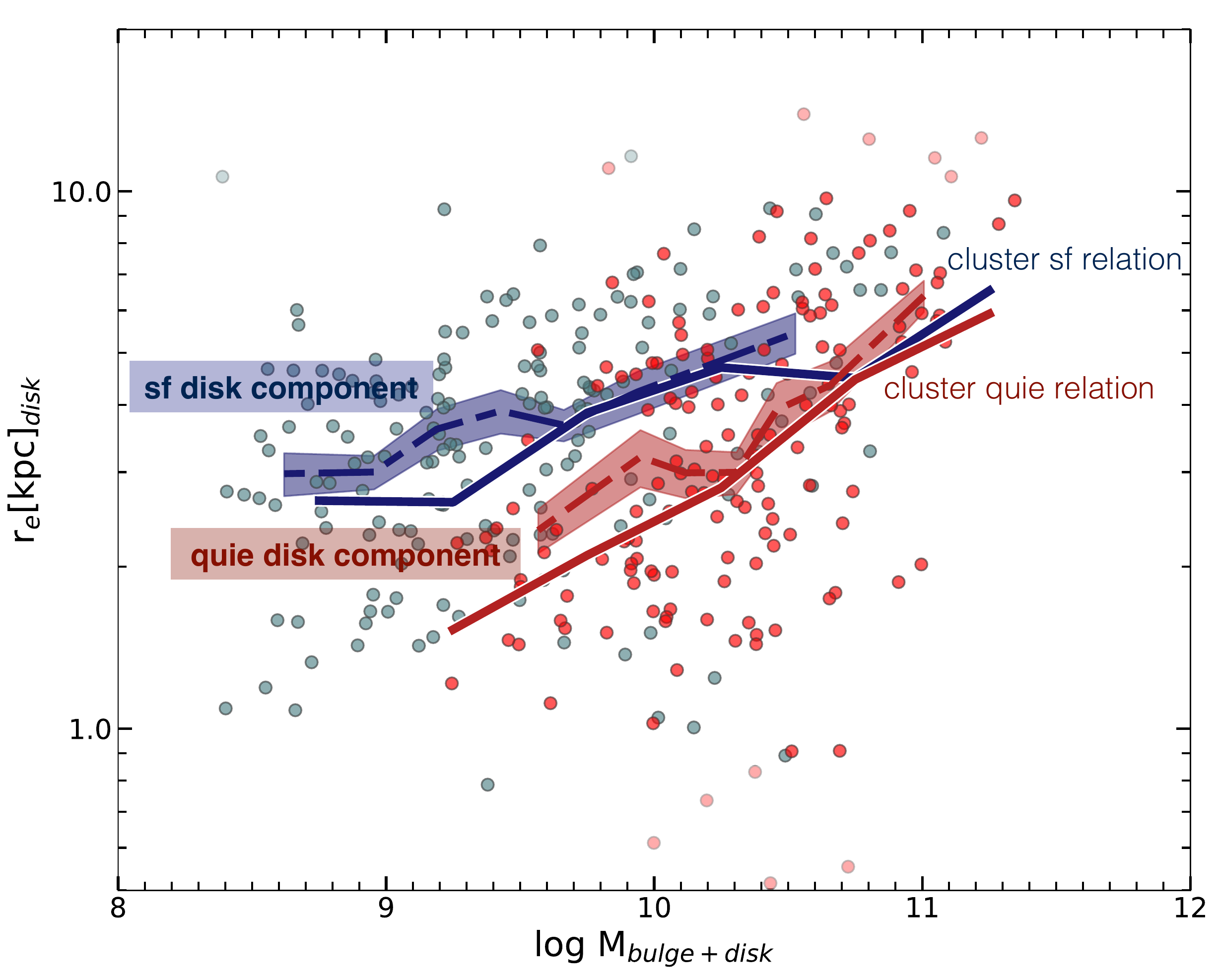}
    \caption{The distribution of sizes of the disk component for 3$\sigma$-clipped data of star-forming (blue) and quiescent (red) galaxies. The overall trends suggest that disk components in quiescent galaxies are smaller than in star-forming galaxies. For reference we also include the whole-galaxy (single component) effective radii relation.} Faint points represent galaxies excluded by the clipping routine.
        \label{fig:MSR_disk_SFQ}%
    \end{figure}

\subsubsection{The disk component of cluster galaxies is smaller for quiescent galaxies}
\label{sec:MSR_disk}

While it is evident that the cluster environment plays an essential role in quenching galaxies, it is less clear how the structure of galaxies change inside the cluster and how these two processes are connected.

Cluster mechanisms that strip away the gas of galaxy halos and disks will dim and redden the galaxies as their stars age. This generally first occurs in disk outskirts, where star-formation rates begin to drop and tidal stripping is most effective \citep{boselli06, bekki09}. 
If  aspect plays a dominant role in actively shaping galaxies in dense environments, we should be able to see this in our data set. 

With our bulge-disk decompositions, we can directly check whether the galactic disk becomes smaller as star-formation ceases in quiescent galaxies. 
This behavior is illustrated in Fig.~ \ref{fig:MSR_disk_SFQ}, where we present 3$\sigma$ clipped fits to sizes for the disk component of star-forming (blue) and quiescent (red) galaxies versus stellar mass of the entire galaxy. The highest B/T quiescent galaxies may be best represented as spheroids only or have unreliable disk measurements. We therefore exclude these galaxies completely. In addition, any failed bulge-to-disk measurements are not used in this fit (see Section \ref{sec:measurements}).
The resulting trends for the cleaned galaxy sample, i.e., the most robust measurements, suggest that the sizes of quiescent galaxy disks are smaller than star-forming disks.   
This may arise from a fading of the outer disk, or brightening of the inner disk. However, we will see later that the cluster environment is associated with increased bulge-fractions (\ref{sec:fractions}), indicating that outer fading is the dominant effect.

In Sec. \ref{sec:comp}, we discussed that the increase of transition objects in clusters leads to smaller averages of sizes of disk-dominated galaxies.
The smaller disk-component sizes of quiescent galaxies cannot on their own account for the measured overall (bulge+disk) sizes of quiescent galaxies. 
However, together with an increase of B/T in quiescent galaxies, resulting from disk fading and possibly some bulge growth, these two effects can account for the dependence of the MSR on star-formation and location inside the cluster.

   \begin{figure*}
   \centering
        \includegraphics[width=\textwidth]{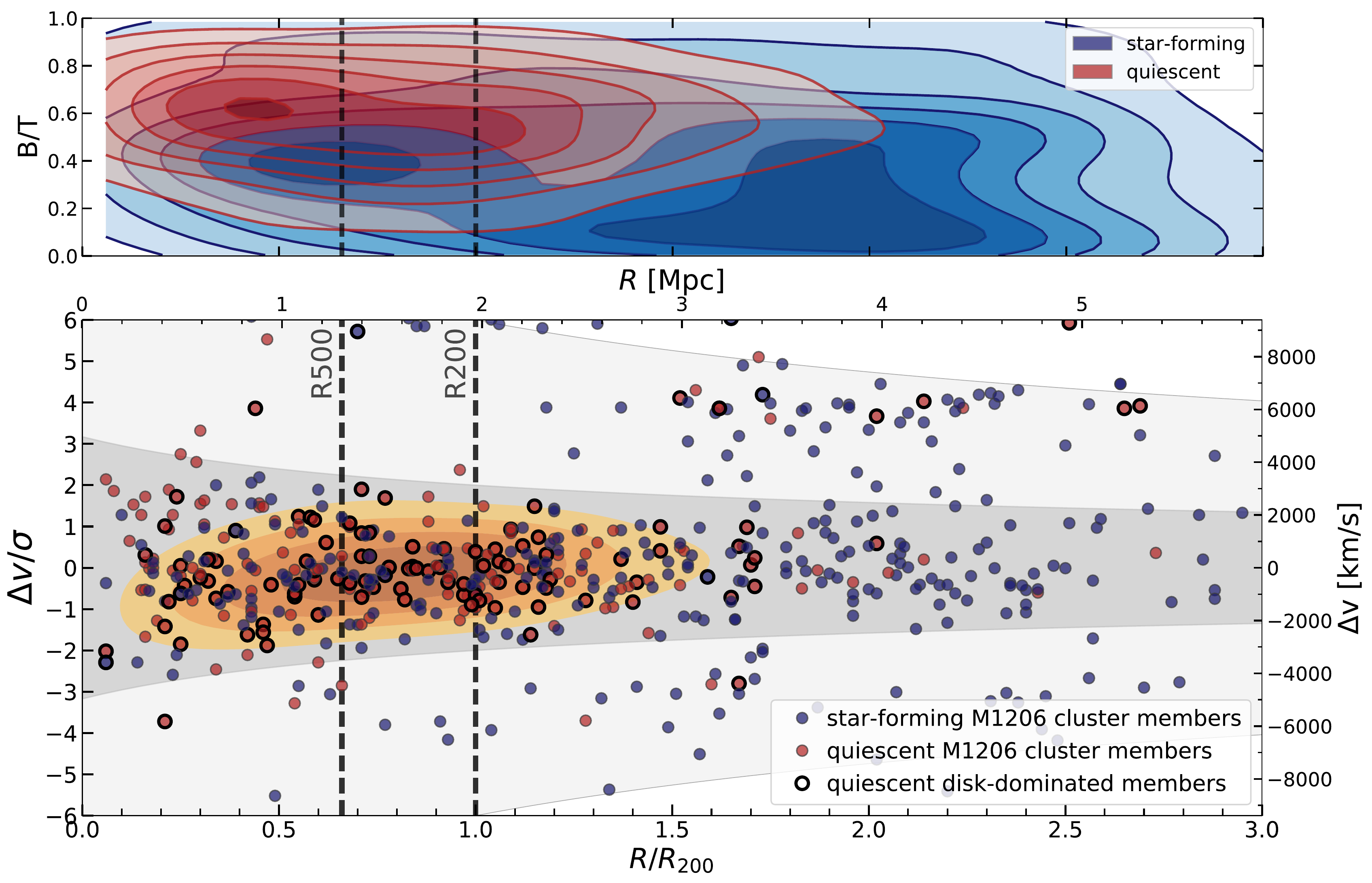}
   \caption{ The main figure shows the distribution of galaxies in phase-space, based on all spectroscopic members of MACS1206 used in this study. We plot radius vs. velocity dispersion of star-forming (blue) and quiescent (red) galaxies. Symbols with black edges show quiescent disk-dominated galaxies, i.e., "red disks". Their density distribution is highlighted with yellow shading, revealing their preferred location between $R_{200}$ and $R_{500}$.
   The dashed vertical line indicates $R_{200}\sim$1.96 Mpc of the cluster (as reported in \citet{biviano13}), and $R_{500}$.
   Galaxies within the 2$\sigma$ lines (inside the grey shaded area) are considered cluster members, between the 2 and 6$\sigma$ lines are infalling galaxies. The top panel shows the 
   2D distribution of all star-forming and quiescent members estimated using Gaussian kernel density estimation (KDE). We  show B/T as a function of cluster-centric radius, normalized by $R_{200}$. The plot demonstrates the distribution of galaxies across the cluster according to their morphology (expressed in B/T): at greatest distances to the cluster center, we find star-forming galaxies with a spread of lower B/T. Their B/T values increase inside $_{200}$. Finally, inside $R_{500}$, quiescent galaxies with higher B/T values prevail.
    }
              \label{fig:SIS_struc}%
    \end{figure*}

\section{Results: The composition and buildup of the galaxy cluster MACS1206}

Our data cover cluster members of MACS1206 out to three virial radii. This means, we sample both the densest part as well as the infall regions of this massive cluster. 
It has been known for decades that morphological Hubble types of galaxies are connected to local density \citep{hubble31} and cluster-centric radius \citep{whitmore93}. In the nearby and intermediate high-z universe, astronomers have observed a significant drop in the fraction of spiral galaxies from 80\% in the field to 60\% in the outskirts of clusters to almost zero in the densest cluster cores. 
The opposite trend is true for early type galaxies \citep{dressler80}. We illustrate this trend for our data by showing the phase-space diagram of the galaxy cluster members (Sec. \ref{sec:radial bins}) and comparing velocity dispersions and fractions of galaxy types as a function of position inside the cluster (Sec. \ref{sec:fractions}).
   \begin{figure*}
   \centering           \includegraphics[width=\textwidth]{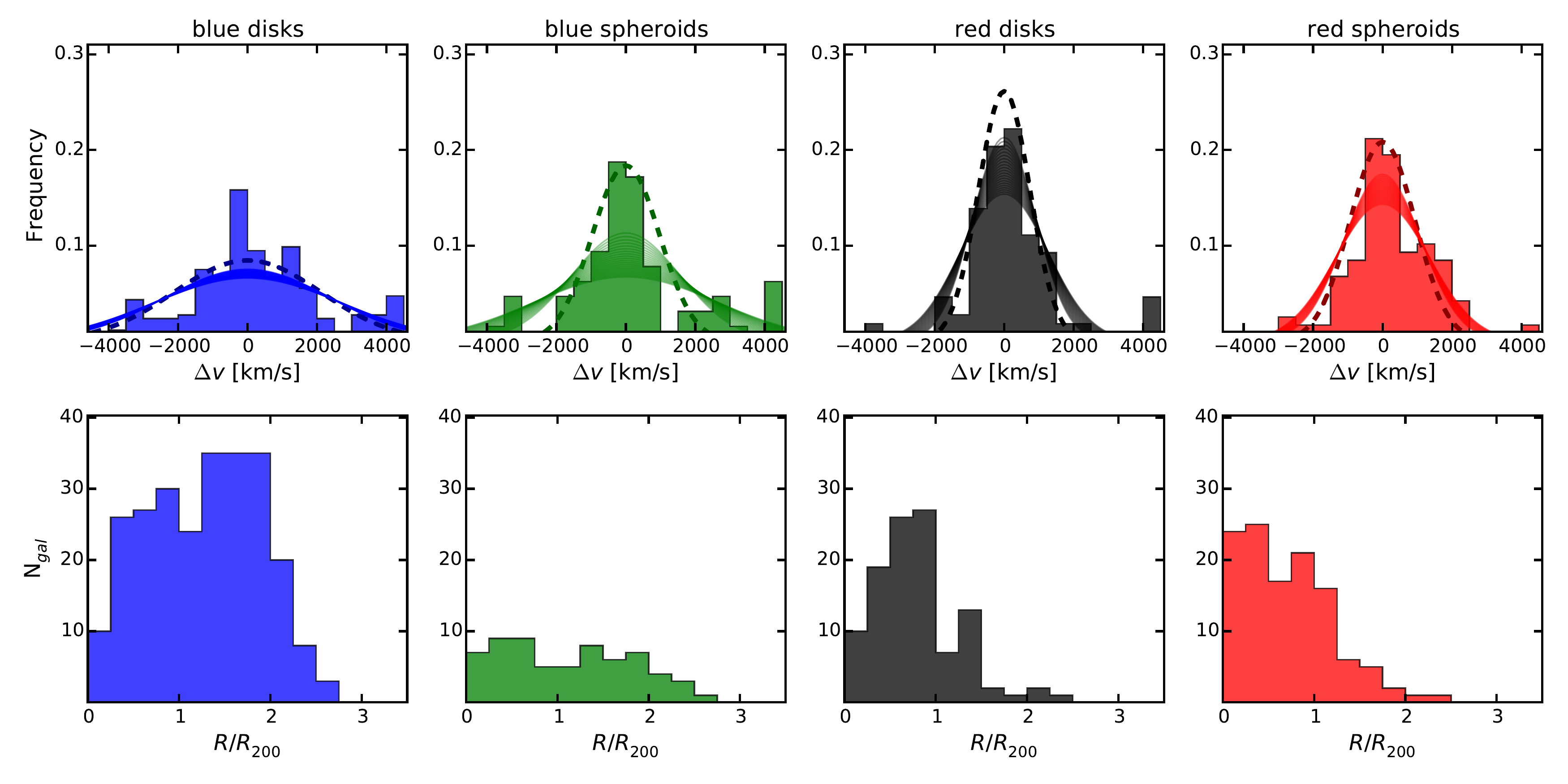} %
   \caption{Velocity dispersions (top) and distribution along cluster-centric radius (bottom) of four galaxy populations inside MACS1206: blue disks (star-forming and $n<2.5$), blue spheroids (star-forming and $n>2.5$), red disks (quiescent and $n<2.5$), and red spheroids (quiescent and $n>2.5$). 
   Blue disks have large velocity dispersions, which indicates a young, unvirialized population. Velocity dispersion of blue spheroids shows two components, a narrow peak and extended wings. This can be a sign for a mix of virialized and unvirialized, possibly infalling, galaxies. 
   Both red sub-populations have small velocity disperions, and are thus virialized. However, they are found in different locations in the cluster.  }
              \label{fig:histos}
    \end{figure*}
\subsection{Phase-space diagram}
\label{sec:radial bins}

Figure \ref{fig:SIS_struc} presents the position-velocity ( phase-space) diagram for star-forming and quiescent members of the massive cluster MACS1206 (with velocity dispersion $\sigma \sim 1087 \mathrm{km/s}$).
The distance to the cluster center, as defined by the brightest cluster galaxy, is plotted against the deviation in velocity space.
While this diagram has some drawbacks, it offers a statistical association of cluster galaxies based on their distribution in the plot \citep{haines12}.

The position-velocity diagram can be used to statistically categorize galaxies into an old, virialized population that was formed locally, or was accreted at an earlier epoch when the young cluster was first assembled. 
These galaxies are found in the cluster core with small line of sight velocities, segregated from the population of galaxies that was accreted at a later stage. These are more common at higher cluster-centric radii and close to the caustics with higher velocities, as they are currently being accelerated into the cluster. 
Galaxies with high velocity dispersion have yet to cross the cluster and thus show how the cluster continues to be constructed.

As galaxies fall into clusters, they are affected by the hostile environment and likely to be quenched in relatively short timescales. 
A combination of intergalactic medium and gravitational interactions are thought to be responsible for galaxy transformations (see, for example, the review by \citet{boselli06} for more detail). Each mechanism acts in a different density regime. 
For this reason, we highlight three physically motivated radial bins in the phase space diagram: 
\begin{itemize}[noitemsep,nolistsep]
\item $R < R_{500}$: the ICM-dominated core
\item $R_{500} < R < R_{200}$: the tidally active region within $R_{200}$
\item $R > R_{200}$: the infall region out to 3 virial radii where galaxies encounter the cluster halo for the first time,
\end{itemize}
where $R_{500}$ ($R_{200}$) denotes the distance where the density is 500 (200) times the ambient density (see Section \ref{sec:data}). 
Our further analysis thus focuses on trends across these radial bins to look for cluster-specific effects on a global scale. 

The top panel of Fig. \ref{fig:SIS_struc} shows the morphology-density relation, summarized as 2D histograms B/T vs. cluster-centric radius.  
It demonstrates how typical B/T ratios shift from lower to higher values with decreasing cluster-centric distances. This is consistent with studies at low redshift and mirrors the known color- and star-formation density relations \citep[e.g.,][]{goto03, blanton05, verdugo08, ziparo14}. 
Trying to understand whether these two effects are related is one of the objectives of this work. 

Dividing galaxies according to their star-formation status, we observe that star-forming galaxies have a larger spread of B/T ratios than quiescent ones. In particular, we do not observe star-forming galaxies with very low B/T values (pure disks) in the inner regions; conversely, they are well represented in the cluster outskirts and infall region outside the virial radius. 
Here, we measure local densities of just a few galaxies per square arcminute, and only 25\% of the galaxies are ellipticals. 
In the transitional region of $R_{500} < R < R_{200}$ we see a large number of galaxies with intermediate B/T values. Many of these are still star-forming; however, an increasingly large fraction of quenched galaxies of slightly higher B/T are present in the same regime. 
In the innermost region $R<R_{500}$, already half of the galaxies are spheroid-dominated galaxies.
They are typically quiescent and their B/T peak around values between $0.6 < B/T < 0.7$. 
A very small number of galaxies with low B/T in the center seem to have not been affected by the cluster, but continue to form stars. 
These galaxies haven’t yet interacted with the cluster or ``survive'' the cluster environment, thus maintaining their star-formation activity.

In the phase-space diagram, we mark ``red disks'' with black outlines and highlight their location with filled density contours in yellow. (As a reminder, this is the transitional population that is quiescent, yet disk-dominated.) 
We find an excess of this population in the intermediate cluster regime, $R_{500}<R<R_{200}$, with a few exceptions that lie in a denser clump in an infalling group.
In a comprehensive study of substructures of MACS1206, \citet{girardi15} have found several more, albeit less prominent, low overdensities in the projected galaxy distribution. 
Apart from these peaks, cluster galaxies classified by their spectral features (in their case into seven sub-classes from passive to very strong emission line galaxies) seem to have a much clearer separation in phase-space than when classified by their star-formation status (from color information) and morphologies like we do in Fig. \ref{fig:SIS_struc}. Emission line galaxies are almost exclusively found outside the virial radius, $R_{200}$, whereas galaxies with prominent disk components also populate areas farther inside the cluster, seemingly remaining their disk structure for longer.

\subsection{Cluster demographics of the galaxy population in MACS1206} 
\label{sec:fractions}

The fraction of galaxies that have been quenched at a given mass is a strong parameter in shaping the average half-light radius and thus the size distribution of galaxies \citep{lilly16}. To understand the stellar-mass -- size relation in MACS1206, it is important to analyze the distribution of the different galaxy populations inside the cluster.

%
\begin{table}
\caption{Velocity dispersions for four subgroups of galaxy members and results of their K-S tests}             
\label{tab:vel_dis}      
\centering                          
\begin{tabular}{l c c}        
\hline\hline                 
Galaxy sub-population & \# & velocity dispersion  \\
 & & (2-$\sigma$ clipped) [km s$^{-1}$]\\
\hline                        
 blue disks & 253 & 2585 (2117) $\pm$178 \\
 blue spheroids & 64 & 2185 (967) $\pm$507 \\
 red disks & 108 & 1096 (741)$\pm$171(741) \\
 red spheroids & 118 & 1247 (932) $\pm$131\\
\hline                                   
\end{tabular}

\begin{tabular}{l c}        
\hline\hline                 
Compared Samples & p-value(\%) K-S test\\    
\hline                        
 blue disks vs. blue spheroids & 6.3\\
 blue spheroids vs. red disks &  <<1 \\
 red disks vs. red spheroids &  <<1 \\
 blue spheroids vs. red spheroids &  <<1 \\
 blue spheroids vs. red disks &  <<1 \\
 red disks vs. red spheroids &  2.4 \\
 \hline                                   
\end{tabular}

\end{table}
   \begin{figure*}
   \centering   
   \includegraphics[width=\textwidth]{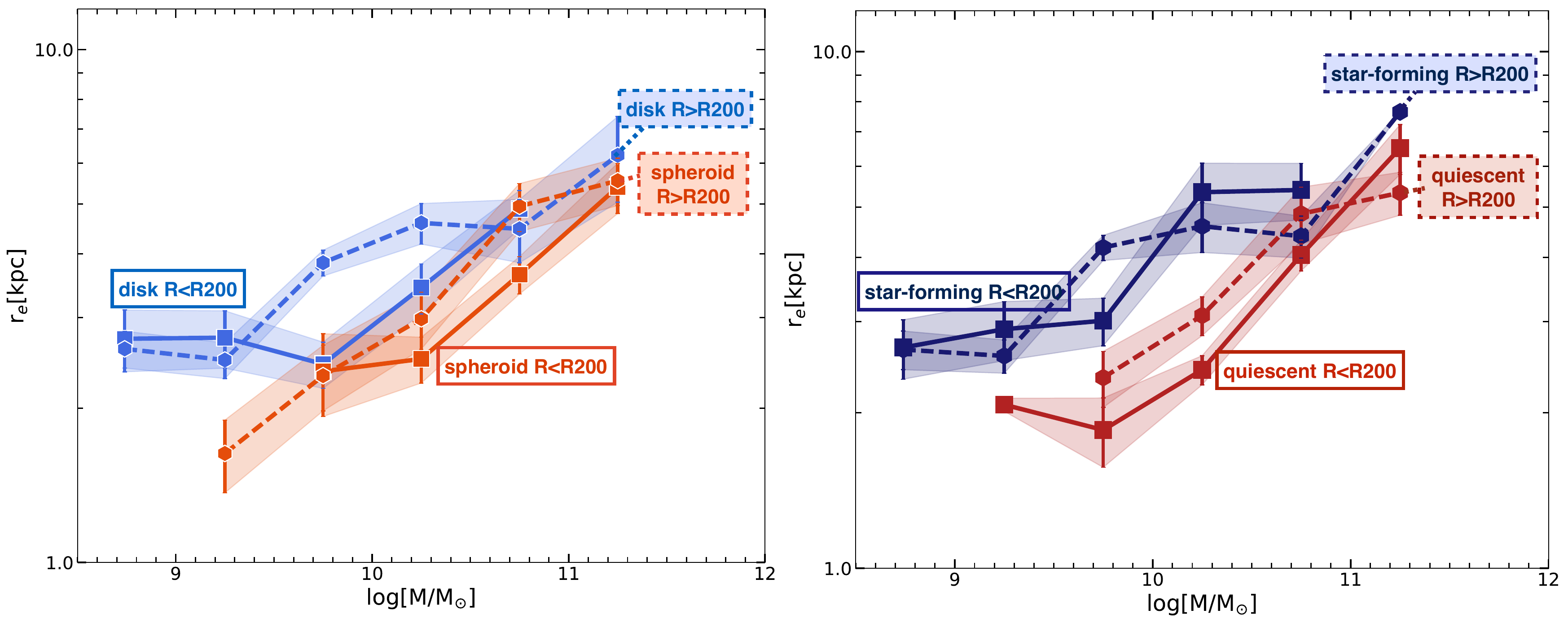} 
   \caption{We plot stellar masses vs sizes in kpc for cluster members inside (solid lines) and outside (dashed lines) $R_{200}$. Left:The galaxies are classified into disk-dominated (blue) and spheroid-dominated (red) galaxies by their Sérsic indices. Disk-dominated galaxies closer to the cluster core are smaller at intermediate masses. Spheroid-dominated galaxies show little size variation in these two regimes. As before, lines represent mean values with their 1$\sigma$ error bands. Right: MSR for star-forming (blue) and quiescent (red) galaxies. Sizes of star-forming galaxies differ marginally and only in one mass bin, whereas quiescent galaxies are consistently larger outside $R_{200}$. This disagreement, as indicated by the 1$\sigma$ error bands, is of the order of 1 to 3 $\sigma$.}
    \label{fig:MSR_struc_sf_2radbins}%
    \end{figure*}
We have seen that MACS1206 hosts a large number of transition objects that are interesting for our galaxy evolution study, e.g., half of the quiescent galaxies are disk-dominated. 
We now use the information given by the phase-space analysis to investigate the individual sub-populations that we introduced in Sec. \ref{sec:comp} and Fig. \ref{fig:MSR_combined}: star-forming disk-dominated ("blue disks"), star-forming spheroid-domianted ("blue spheroids"), quiescent disk-dominated ("red disks") and quiescent spheroid-domianted ("red spheroids").
The top row in Fig. \ref{fig:histos} shows the distributions of velocity dispersions of these four groups, where bands show the variations of normalized Gaussian fits from a bootstrapping procedure of 100 resamplings, indicated by the thin lines. Each panel shows the fractions of the considered population.
We tested whether the four samples differ from a normal distribution by applying a normaltest \citep{dagostino73}. In general, the populations are well represented by Gaussians. 
An exception are blue disks: their p-value of the normaltest reject that they are drawn from a normal distribution.

The values of their velocity dispersions are reported in Table \ref{tab:vel_dis}, along with results from the Kolmogorov-Smirnov (K-S) test\footnote{The K-S test assesses the statistical significance of the difference of populations by comparing the shapes of their cumulative distribution functions.} of the radial distribution in the cluster (shown in the the bottom row of Fig. \ref{fig:histos}).
The velocity dispersion of the two star-forming subgroups are similarly high (2585 km\,s$^{-1}$ for blue disks and 2185 km\,s$^{-1}$ for blue spheroids), indicating a new, unvirialized population of the cluster. 
However, they are the only pair where the K-S test does not rule out the null hypothesis (that both populations are drawn from the same parent distribution), with a probability of 6.3\%.
Blue spheroids seem to have two components of the velocity dispersion profiles: a narrow central distribution, very similar to the red galaxies and extended wings that bias the velocity dispersion and are responsible for their overall wide distribution and high errors. The green dashed line in Fig. \ref{fig:histos} indicates this narrow distribution (with a velocity dispersion of 1078 km\,s$^{-1}$) after a 2$\sigma$-clipping was performed. Sigma-clipping of the other populations, indicated by dashed lines and written in parentheses in table \ref{tab:vel_dis} did not significantly change their velocity dispersion.
This might indicate that most of these galaxies form part of a virialized population, similar to the quiescent galaxies. Alternatively, these could be galaxies of the inner cluster regions that still actively form stars.
Blue spheroids are located in the cluster center as well as in the outskirts. Whether this population has a relatively high fraction of infalling galaxies, or whether what we see is simply noise, is unclear. Since only 10\% of our cluster members are blue spheroids (see Fig. \ref{fig:MSR_combined}), the numbers available for this group of galaxies is very low.

Relating the two sub-types of quiescent galaxies, we see that they have comparably small velocity dispersions, regardless of their morphological type (red disks and red spheroids). 
However, their spatial distributions differ: red disks avoid the innermost regions of the cluster, an area where red spheroids dominate. A reason for this could be a difference in age. We expect the location of an older, virialized population to be in the cluster center and a younger population farther outside \citep{haines13}. 
\break

The relative number of galaxies of different populations depends on the density of the environment. 
This is reflected by the changing fractions of star-forming and quiescent galaxies along the cluster-centric radius (see bottom row of Fig. \ref{fig:histos} for changing fractions of the four sub-groups).

The number of star-forming galaxies that make up 72\% of the population in the outskirts of the cluster, decreases dramatically inside the virial radius $R_{200}$, reducing to 40\% of the population closest to the cluster center. These fractions are consistent with studies of the Butcher-Oemler effect at similar redshifts \citep{ellingson01, hennig17}.

Consequently, the fraction of quiescent galaxies in each regime more than doubles (from 28\% of galaxies at $R>R_{200}$ to 60\% near the cluster center). 
In MACS1206, around half of the quiescent galaxies are disky. While 60\% of quiescent galaxies in the intermediate regime have disks, this drops to 33\% inside $R_{500}$. In addition, as we have discussed, red disks are absent from areas closest to the cluster center.

\subsection{Do galaxy sizes depend on cluster-centric radii?} 
\label{sec:MSR_radial bins}

In this chapter, we analyze variations of the stellar-mass--size relation inside and outside $R_{200}$.
The aim here is to investigate any transition signatures with our observations of cluster galaxies in MACS1206 and relate them to galaxy sizes. We follow the MSR of star-forming/quiescent, $n<2.5$/$n>2.5$ and over different cluster radial bins.
We pay particular attention to a transitional population of quiescent disk-dominated galaxies that influence the behavior of the relation inside the cluster before turning to our bulge-to-total measurements.

In the previous sections, we showed that galaxy sizes inside a cluster depend on their type. 
We have also seen that the makeup of galaxies inside a cluster changes considerably over the three radial bins probed in this paper, and that the correlation between morphology and star-formation activity is not absolute.

Next, we therefore examine the mass-size relation for $n<2.5$ disk-dominated and $n>2.5$ spheroid-dominated galaxies (left panel in Fig. \ref{fig:MSR_struc_sf_2radbins}), and star-forming and quiescent galaxies (right panel in Fig. \ref{fig:MSR_struc_sf_2radbins}) as a function of cluster-centric radii. In order to increase our number statistics, we do this for galaxies inside and outside $R_{200}$, i.e., in regions for which we might expect diverging properties. 

While the direction of trends towards smaller sizes closer to the cluster center is the same for star-forming and disk-dominated galaxies (just as they are for quiescent and spheroid-dominated), changes in size are more noticeable in the structures of galaxies. 
At intermediate masses, galaxies with single Sersic $n<2.5$ are smaller closer to the cluster center than they are farther out. The error bands show that this discrepancy is significant (between 2 and 6 $\sigma$) at mass ranges where disk galaxies are abundant in both radial bins.
Spheroid-dominated galaxies tend to be smaller inside the cluster center. However, larger errors, due to the large scatter, reduce the significance for this population.

In the right panel of Fig. \ref{fig:MSR_struc_sf_2radbins} we compare sizes for star-forming and quiescent galaxies inside and outside $R_{200}$. While we do not see any conclusive differences for galaxies that retain their star-formation activity in the cluster center in comparison to star-forming galaxies at $R>R_{200}$ (except in one mass bin), quiescent galaxies show a tendency (2 $\sigma$) to be smaller inside $R_{200}$. 

\begin{figure}
   \centering        
   \includegraphics[width=\columnwidth]{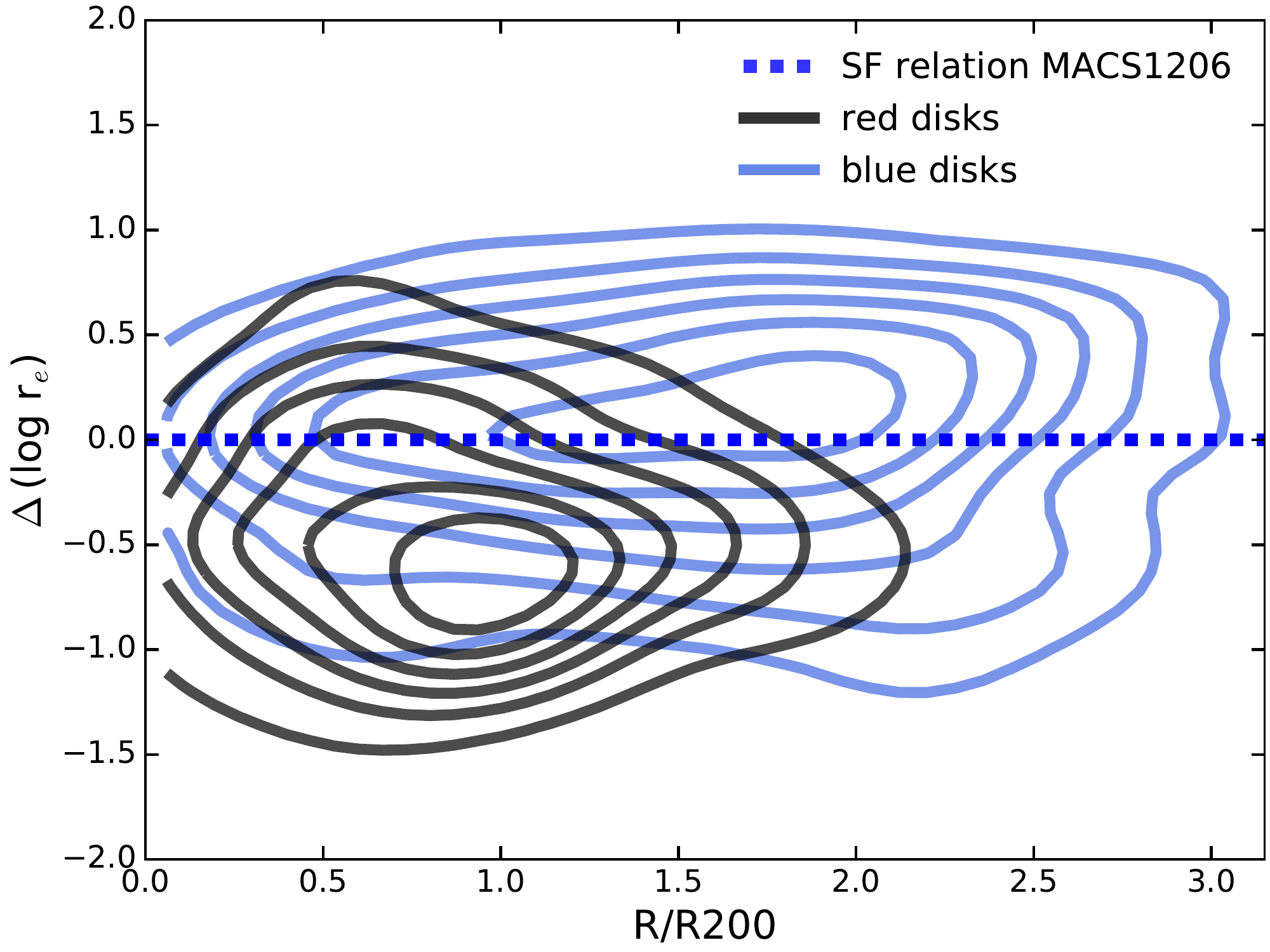} 
   \caption{Differences in size to the star-formation relation (as shown in Fig. \ref{fig:MSR_hubble}) vs. cluster-centric distance, normalized by $R_{200}$. We show density contours for blue and red disk galaxies. At $R_{200}$, an increasingly important population of "red disk" galaxies is responsible for a decrease of sizes of disk-dominated galaxies.}
    \label{fig:MSR_bluered}%
    \end{figure}

For an explanation of these signatures we need to combine our observations: 
sizes for disk-dominated galaxies decrease inside $R_{200}$, because half of them have already quenched. Already in Fig. \ref{fig:MSR_combined} we have seen that quenched disks are much smaller than star-forming disks. Fig. \ref{fig:MSR_bluered} now clearly demonstrates the rise inside $R_{200}$ and decreased sizes of this population by contrasting the size- and spatial distribution of star-forming (in blue) and quiescent (in black) disks inside MACS1206. Evidently, the appearance of “red disks” in the intermediate region, and their starkly smaller sizes are responsible for pulling down the MSR of disks inside $R_{200}$.

Red disks can also explain the difference of the MSR of quiescent galaxies inside and outside $R_{200}$. More than half of all red galaxies outside $R_{200}$ are disk-dominated.
Their sizes are slightly larger outside than inside $R_{200}$, presumably because they have recently quenched and their disks have not had as much time to fade. This is supported by the finding of an increasing fraction of galaxies with higher B/T closer to the cluster center (Fig. \ref{fig:SIS_struc}) and their observed smaller sizes (Fig. \ref{fig:MSR_BT}).

Combining our observational results, we conclude that the location of a galaxy in the mass-size relation mainly depends on their structure/bulge fraction (which is connected to a reduction of the disk component) and their star-formation status:
galaxies have increasingly smaller effective radii at increasingly larger B/T and are more likely to be quiescent. 
Any representation of the mass-size relation will therefore be influenced by the varying fraction of galaxies corresponding to the different types, i.e., the composition of the cluster.
However, while the relative number of galaxies shift to higher B/T values at smaller cluster-centric radii, the sizes of galaxies at a specific B/T largely stays the same.
To put in another way, at fixed morphology (B/T), we observe a strong variation in the number of star-forming and quiescent galaxies with cluster-centric radius.
Once again we come to the understanding that the sizes of galaxies in clusters vary as a result to a combination of several effects that are almost impossible to disentangle.  

\section{Discussion} 
\label{sec:discussion}

In this paper, we have used the stellar-mass—size relation as a tool to 
investigate how a massive galaxy cluster affects its members.
This revealed variations of the galaxy size distribution that relate to changes of the star-formation activity and morphologies of galaxies in clusters. 

There are several mechanisms specific to clusters that are capable of suppressing star-formation by removing or disrupting the interstellar gas in galactic disks and halos.
Firstly, because the extended halo gas component is only loosely bound to the galaxy, it is easily stripped in cluster conditions. 
Numerical simulations show that after entering a galaxy cluster halo, the tidal field of the cluster first distorts and finally disperses the disk halo gas into the intra-cluster region. The disk does not receive fresh gas, and feedback from active galactic nuclei and supernovae becomes more efficient in ejecting disk gas. As a result, star-formation ceases, and the galaxy becomes more fragile \citep{bekki01}.
These ``strangulation'' processes \citep{larson80} are characterized by the immediate onset of a gradual and slow decline of the star-formation \citep[$\tau > 1\ \mathrm{Gyrs}$,][]{mcgee09, weinmann10,delucia12}. 
Eventually, cluster galaxies use up their gas supply and become passive. While the processes acting on the halo gas are important, they cannot be responsible for any rapid changes and are not linked to a specific location in the cluster.

As galaxies continue to fall into denser regions of the cluster, they are susceptible to processes exerted by the hot intracluster medium acting on the gas in the disk as well as the halo. 
The gas component in the galaxy experiences ram pressure that, if sufficiently strong, strips the disk gas \citep{gunngott72}. Because the surface densities of stars and gas declines as a function of galacto-centric distance, stripping will be more effective in the outskirts of galactic disks \citep{abadi99}.
The suppression of star-formation thus starts in the outskirts such that the galaxy quenches in an outside-in fashion \citep{schaefer17}.
This is likely a rapid process, starting around the virial radius of the cluster that leads to a fast quenching of star-formation.
Especially if the gas disk is already porous, ram-pressure stripping may in some cases strip the entire cold gas reservoir within $\sim 10^8 \mathrm{yrs}$ \citep{quilis00}.

Furthermore, the gas-poor galaxies are more easily disrupted by gravitational interactions acting on the stellar component. These include minor-mergers \citep{hopkins09} and galaxy ``harassment'': the effect of rapid close encounters of cluster galaxies in combination with interactions with the cluster potential itself. As a consequence, the fragile disks are impulsively heated and tidally stripped \citep{farouki81, moore96}. 
At the same time, these high speed encounters can also cause gas to sink towards the center of the disk, which helps in the formation of the bulge and further consumes gas.
While this process likely requires long timescales to become effective, simulations of harassment are capable of reproducing observed properties of environmentally transformed early types in clusters \citep{mastropietro05}.

All of these processes have profound implications on the morphologies of galaxies. Halo- and disk-gas stripping are mechanisms that eventually stop the galaxy from forming stars.
As a consequence of the gas stripping in the outskirts first, the outer regions of the stellar disks fade and the spheroidal component becomes more dominant. Because the shape of the light profiles change, they are best fit with higher S\'ersic index parameters $n$. 
The absence of star-forming regions also makes passive galaxies appear smoother.
In addition, tidal processes may re-arrange the distribution of the stars, which further affects their morphologies.

Previous studies have shown direct observational evidence of cluster mechanisms affecting the gas and stars in galaxies. 
Surveys in nearby clusters have shown that up to 50\% of observed late-type galaxies show tails of ionized gas (sometimes with embedded star-forming regions) extending from the disks and pointing in the opposite direction of the cluster center \citep{yagi10, boselli14}.
These observations are taken as direct evidence of ongoing ram-pressure stripping events.

Observations of an abundant population of passive disk-dominated galaxies in clusters offer another clear signature for environmental processes. In the past, publications have highlighted the class of anaemic \citep{vandenbergh76} or passive spirals \citep{goto03}, with findings of quenched HI-deficient disks \citep{vogt04}. 
Accordingly, \citet{koopmann04} compared $H\alpha$ observations with optical data and found a high fraction of $H\alpha$-truncated galaxies in the cluster environment. The galaxies, though affected by stripping from the ICM, have relatively undisturbed stellar disks with star-formation confined to the inner disk.
The high abundance of these galaxies suggests that cluster galaxies experience a common phase of ``red disks'' on their way to the red sequence  \citep{moran07, bundy10}. 
The identification of these objects highlight a transition phase in the evolutionary sequence of galaxies influenced by cluster mechanisms. 

Throughout this paper, we presented clear signatures linked to the well defined observables colors, sizes, masses and bulge-to-total ratios.
Disk- and spheroid-dominated galaxies were separated by their single S\'ersic index $n<2.5$ and $n>2.5$ respectively and star-forming and quiescent galaxies were defined by their BRI-color selection.

We first saw that the stellar-mass--size relation of star-forming and quiescent galaxies in the massive cluster MACS1206 roughly resembles that of field galaxies, with two exceptions:  high-mass star-forming galaxies (above $10^{10}\mathrm{M_{\odot}}$) are smaller, and quiescent cluster galaxies are slightly larger than their field counterparts.
While the time for galaxies to cross the cluster is independent of their mass, massive star-forming galaxies undergo a slow quenching process as they are accreted onto the cluster, consistent with a gradual shut down of star-formation \citep{haines13}.  The galaxies do not fully quench at first crossing and are only partially affected by the time they reach the turning-point.
As we discussed above, the cluster mechanisms first shut off star-formation in the outskirts of the galaxy, fading the outer disk.
Because high mass galaxies are typically large, even after the disk outskirts are stripped of gas, enough of the disk component remains to form stars. In other words, their truncation radii will not reach all the way into the inner regions. 
Rather, their disk stars fade slowly, gradually reducing their sizes while still continuing with a low level of central star-formation.

Low mass galaxies are more vulnerable to the environment and generally quench and change their morphology more quickly. According to the MSR, low mass galaxies are smaller, so the effect of ram-pressure has the possibility to strip the star-forming disk gas completely in one crossing time. Quenching therefore occurs so fast that they move out of the star-forming category immediately and become part of the quiescent sample.
The different quenching timescales for low- and high-mass galaxies thus explain the cause of the observed differences between the MSRs of cluster and field galaxies.

We then moved on to compare the mass-size relation of galaxies in MACS1206 divided by their star-formation status with those selected by structure.  
We saw that at intermediate masses, disk-dominated galaxies are on average slightly smaller than star-forming galaxies. We attribute this to a population of smaller ``red disks'' that are particularly prevalent in clusters (22\% of our entire sample and half of all quiescent galaxies). 
This ``red disk'' population is classified as quiescent in one selection and disk-dominated in the other, which leads to a larger number of disks than star-forming galaxies and occur at all masses above 10$^9 M_{\odot}$ inside the cluster. 
Later, when we identified the cluster radius at which galaxy signatures change, we were able to link the intermediate cluster regime, between $R_{500}$ and $R_{200}$, to the rise of this population.
It is in this region where galaxies become subject to an increasing dominance of the ICM and the general dichotomy of galaxies into early and late types breaks down.
Dynamically, quiescent disks are virialized, just as the quiescent spheroid-dominated population they are most likely going to evolve into. However, red spheroids populate a different location in the cluster; they dominate in the core where the oldest members of the cluster reside - a region red disks avoid.
Inside $R_{200}$, the sizes of red disks are comparable to the sizes of red spheroids, however, without a complete compaction of the galaxies's stellar distributions.

We link the smaller sizes to two signatures we observe: bulges becoming more dominant and disk sizes decrease. 
This suggests that by the time the galaxies enter the "red disk" phase, a fading of the outer disks and morphological transformation are on its way.  
It is appealing to assume that the quenching of star-forming galaxies also depend on their internal structure and that environmental quenching is directly related to bulge growth.
Bulges are thus either prerequisite for the transformation of star-forming galaxies to the red sequence or are formed (or grow) by this quenching process \citep{dominguez09, capellari13}.

Because the disk structures are kept in tact, a gentle quenching process is likely responsible for shutting off the star formation in red disk galaxies.
The observed signatures may be the consequence of the removal of gas, most notably by the fast-acting ram-pressure stripping that operates in the intermediate regime between $R_{500}$ and $R_{200}$. While this is a likely candidate to rapidly strip the galaxy of its star-forming outskirts, fading of the outer disks may be at least partially caused by harassment and begun by strangulation. 
These are processes less violent than e.g., major mergers that lead to a destruction of the disk.
In this scenario, galaxies closer to the core have typically had more time for their disk component to fade, reflected in a lower mass-size relations for disk galaxies inside $R_{200}$.

Although ``red disks'' do occur in the field ($\sim 10\%$ of galaxies in low density environments), there the strong preference for higher masses suggests a different cause. 
Most field ``red disks'', can be attributed to edge-on inclination and significant dust reddening effects \citep{bamford09}. 
In clusters, their fraction is around five times higher, supporting our interpretation that the ``red disk'' population we observe in MACS1206 results from processes caused by the cluster environment. 

\citet{wolf05}, using COMBO-17 cluster data, found that, while optically quiescent red spirals continue to form stars in the inner dust obscured region after their outer disks have been stripped by the ICM. 
In this scenario, star-formation is hidden from the optical light, and only visible in the infrared. The red colors are then partially the result of the high dust extinction.
This central star-formation increases bulge sizes which fits well with our measurements of increasing bulge-to-total values.
In their sample of super-cluster galaxies, dusty spirals prefer medium-density regions, similar to our ``red disk'' sample.

The global mass-size relations for quiescent and spheroid-dominated galaxies in MACS1206 are comparable to each other. 
The early-type population is likely dominated by galaxies that have quenched a long time ago, and we might therefore expect them to be (uniformly) compact.
However, they have been subject to cluster specific effects like harassment and - to some extend - dry minor merging that are capable of puffing up the outer parts of galaxies. In addition, a population of more extended, newly quenched galaxies is added to the mix.
While the high mass end of the quiescent population is globally in place at this redshift ($z\sim 0.5$), galaxies at lower masses thus evolve inside the cluster.

Quiescent galaxies in MACS1206 are not uniformly small at all cluster-centric radii. They are smallest inside $R_{200}$, where clusters host a high fraction of galaxies that have quenched and become compact at high redshifts \citep{vanderwel14, lilly16}. They have been part of the cluster the longest \citep{haines13}. These galaxies boast red colors, high masses, high B/T and relatively small effective radii. This is consistent with the findings of a systematic survey of the COSMOS field by \citet{damjanov15} that highlight a preference of massive compact galaxies for denser regions.
However, red galaxies do not retain their compactness completely everywhere in the cluster. They are intermixed at increasingly large cluster-centric radii with newly quenched (and slightly larger) galaxies whose gas has been quickly removed by the cluster environment.
After being stripped of their gas, lower mass star-forming galaxies quench and change their morphology quickly.
As a consequence, their S\'ersic indices increase and the galaxies move from the disk- to the spheroid-dominated category of our sample. 
As their remaining disks continue to fade, the newly quenched galaxies have larger sizes outside than inside $R_{200}$.

The mass-size relation can therefore be understood as the consequence of a mix of different progenitors formed at a range of quenching epochs and by a number of quenching mechanisms. As a consequence, the galaxy population inside a cluster is the reflection of a complex combination of effects. 
This in turn weakens any signal of specific  environmental effects for the cluster population.

\section{Summary}

The stellar-mass -- size relation can be used to improve our understanding of galaxy evolution in varying environments.
We analyzed the size distribution of cluster members of the massive galaxy cluster MACS J1206.2-0847 first in comparison to the field and then for the entire cluster and at decreasing cluster-centric radii. 
Any dependence of the environment on the size distribution of galaxies is mass and morphology dependent.
We therefore follow trends for galaxies in $0.5$dex bins in mass, divided into star-forming and quiescent galaxies, into disk- and spheroid-dominated galaxies and as a function of B/T. 
It becomes evident that changes in star-formation act on different timescales and in different environmental regimes, and are not necessarily directly related to changes in morphology.

Our investigation revealed the following observational results and correlations:

\begin{itemize}
\item In comparison to the field, we measure smaller sizes for massive late type galaxies and larger sizes for early type galaxies in the cluster. The first, we attribute to the difference in transformation timescales for high and low mass galaxies: essentially, the mass of a galaxy influences its resistance against violent cluster-specific effects. The more massive star-forming (disk) galaxies survive inside the cluster for longer and therefore also keep forming stars for longer (i.e., they continue to be part of our star-formation selection), while their outer disk slowly fades and, as a result, smaller effective radii are measured.
The second, we explain by showing an increasingly important population of newly quenched galaxies with larger sizes, many of which are transitional objects. 
 
\item At intermediate masses, the sizes of disk-dominated galaxies tend to be smaller than sizes of star-forming galaxies (classified by their color). 
This is due to the rising number of transitional objects, most notably galaxies that are quiescent but have retained their disk morphology, in the tidally active intermediate and inner cluster regime $R<R_{200}$. Here, ``red disk'' galaxies resemble the size distribution of red spheroid galaxies, but not their stellar structure or bulge fraction.
Not only are they responsible for a decrease of the overall sizes for disk galaxies, they contribute to larger average sizes of quiescent outside $R_{200}$.

\item The investigation of the spatial distribution and velocity dispersions of different subclasses (i.e., blue disks, blue spheroids, red disks, red spheroids) revealed that red disks are a virialized population; however, their location in the cluster suggests they are younger members than red spheroids. 
The unvirialized blue disks are preferentially located at higher cluster-centric radii while star-forming spheroids are made up of a mix of populations spread out over the projected area of the cluster.

\item Inside the cluster, the stellar-mass -- size relation is not a representation of two distinguishable trends, one for star-forming disks and one for quiescent spheroids. It rather presents itself as a picture of galaxy sizes smoothly decreasing as a function of bulge fraction with little mass variation.
Independent of their B/T and mass, quiescent galaxies are smaller than their star-forming counterparts. 
This is due to an outer disk-fading and possible bulge growth that accompanies the varying fraction of star-forming and quiescent galaxies already present in the cluster. 
We make a combination of ram-pressure stripping of the cold gas and other forms of gentle, more gradual gas starvation responsible for this newly quenched population, dominated by galaxies with high bulge fractions.
\end{itemize}

While this study focused on the environmental dependencies, we anticipate to continue this investigation with multi-band high resolution images of cluster core galaxies to further study the colors of the galaxy components. We need to confirm the significance of individual populations like transition objects and expand this study with an inspection of the stellar population of cluster members. While disentangling the formation and evolution mechanisms of galaxies inside clusters is a complex endeavor, the prospects are increasingly auspicious with high quality data.  
\\
\\
\\
\begin{small}
\textit{Acknowledgements} This work benefited from the input of the CLASH and CLASH-VLT collaboration, who not only prepared much of the imaging and spectroscopic data, but also provided valuable comments and ideas to the undertaking of the research summarized here. 
We would like to extend our thanks to the referee for reading our manuscript carefully and offering useful suggestions and comments, as well as to Alyssa Goodman who gave some excellent advice on how to visualize complex information. 
We acknowledge the crucial contribution of the ESO staff for their involvement in operations and support. This study is partially based on data collected at Subaru Telescope, which is operated by the National Astronomical Observatory of Japan. The CLASH Multi-Cycle Treasury Program is based on observations made with the NASA/ESA Hubble Space Telescope.
UK acknowledges the support from the University of Vienna and the Marietta Blau Grant, financed by the the Austrian Science Ministry as well as support from the ESO visitor programme. 
She thanks the University of Nottingham for hosting her while working on this paper. This publication is supported by the Austrian Science Fund (FWF).
\end{small}

\bibliographystyle{aa} 
\bibliography{bibliography} 

\begin{appendices}
\section{Appendix A: Investigation of the completeness of the presented spectroscopic sample}
\label{sec:appendix}
   \begin{figure}[!ht]
   \centering
        \includegraphics[width=0.95\columnwidth]{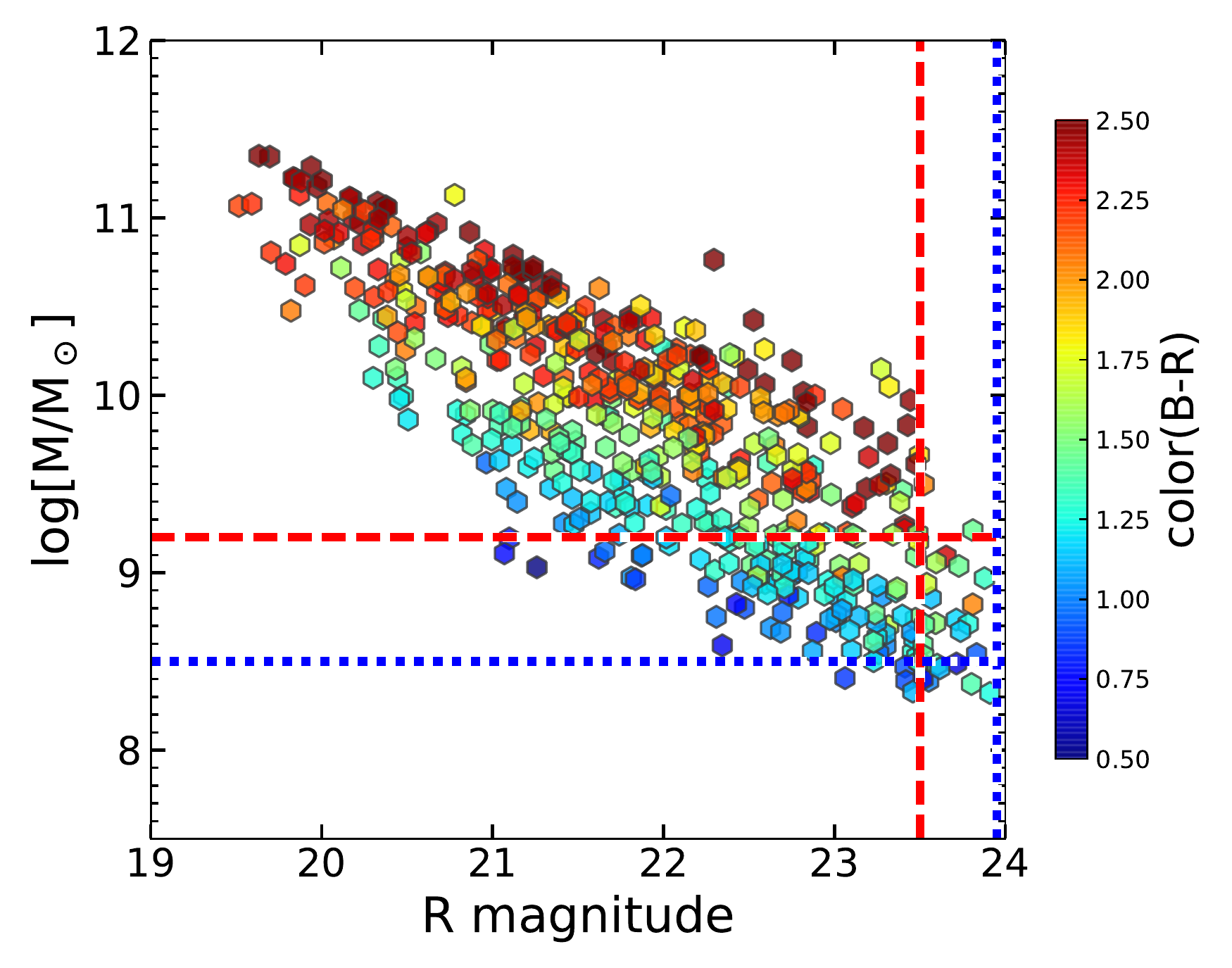}
   \caption{Stellar masses as a function of R-band magnitudes for our sample of MACS1206 members. Galaxies are color coded by their (B-R) color. The relation between magnitudes and masses allows the translation of completeness magnitudes to mass limits used in our study, represented by dashed (for quiescent galaxies) and dotted lines (for star-forming galaxies).
   }
              \label{fig:R_mass_color}%
    \end{figure}
In Figure \ref{fig:R_mass_color}, we demonstrate our mass completeness limits from a conversion of R-band apparent magnitude to masses (in log[M/M$_{\odot}$]), based on their relation. 
Intrinsically, star-forming galaxies are less massive than quiescent galaxies, therefore we adopt color-dependent mass limits, conform with our comparison field sample. 
The vertical dashed/dotted lines represent the completeness magnitude of our sample for red/blue galaxies, and the horizontal line represents the corresponding completeness mass.
For (color-selected) star-forming galaxies this results in a completeness mass limit of $\log(\mathrm{M/M}_{\odot}$)=8.5 (blue dotted line); for quiescent galaxies our limit is $log(\mathrm{M/M}_{\odot}$)=9.2 (red dashed line).

\bigbreak
Next, we investigate possible incompleteness in our spectroscopic sample to ensure that it provides an unbiased representation of the galaxy population.

\begin{figure}[!hb]
   \centering
        \includegraphics[width=\textwidth]{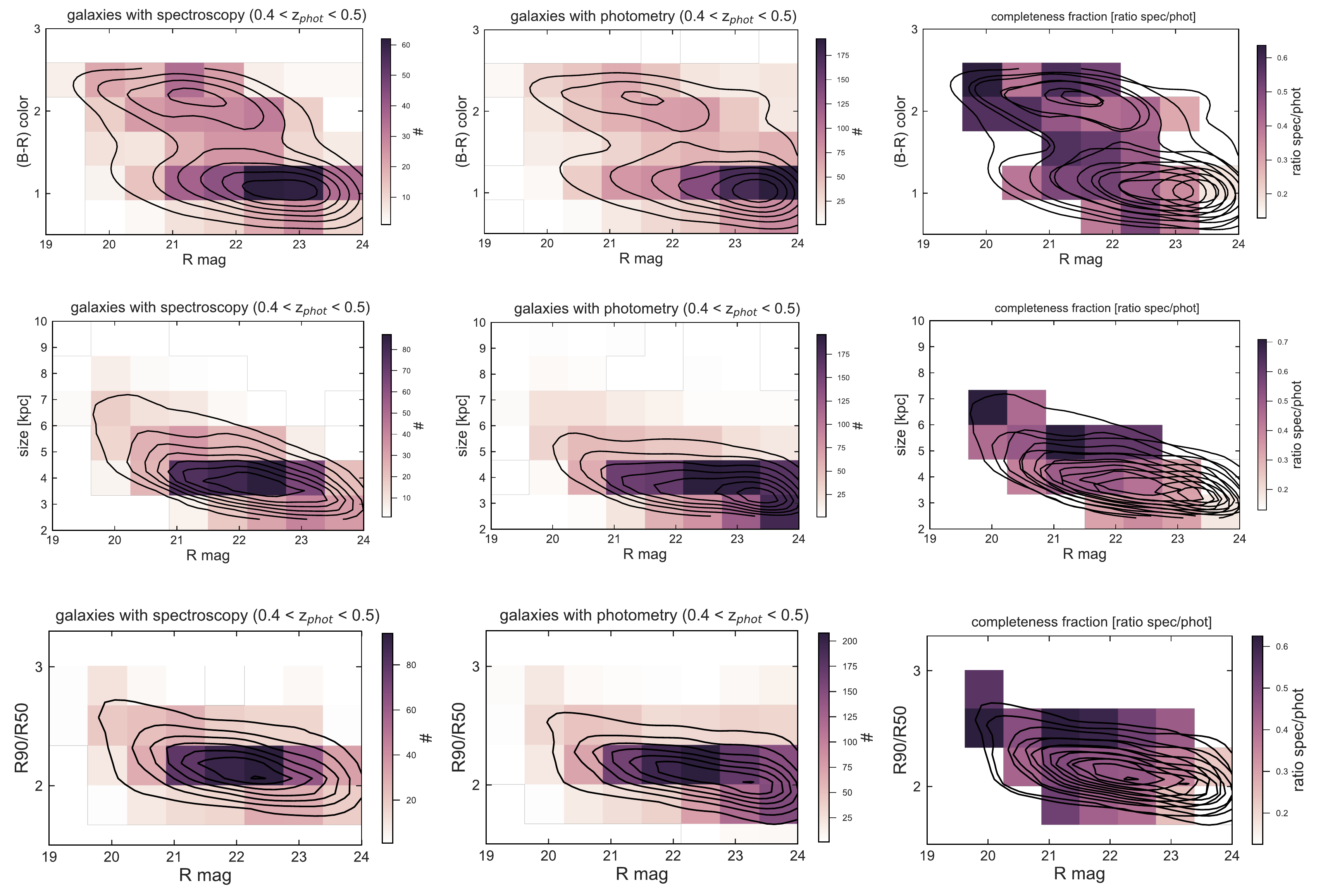}
   \caption{2D histograms of color (first row), size (second row) and concentration (third row) as a function of apparent magnitude. Black contours highlight their Gaussian kernel density estimation. The first column, shows the distribution of galaxies in the photometric redshift range 0.4<$z_{phot}$<0.5, calculated from Subaru SuprimeCam, and matched to the field-of-view of VIMOS. The middle column shows the same for galaxies with spectra. The right column demonstrates completeness fractions in bins with at least 10 galaxies. The two samples hardly differ in the y direction. We correct for variations in the magnitudes with weights.}
              \label{fig:2dhisto_all}%
    \end{figure}

Incompleteness may depend not only on observed magnitude (and mass), but also on other measured parameters, like galaxy colors, sizes and concentration.
In Fig. \ref{fig:2dhisto_all} we look at the on-sky completeness of the sample in terms of color, size (from FLUX\_RADIUS measurements) and R90/R50 (the ratio of galaxy radii containing 90\% and 50\% of the flux) as a measure of concentration as a function of magnitude. Figures in the first row show 2d histograms of galaxies in the photometric sample. This was compiled from a SExtractor run, cleaned of stars and matched to the area of the VIMOS field of view to overlap the spectroscopic sample. Furthermore, we selected the galaxies in a photometric redshift range between z=0.4 and z=0.5 to roughly frame the galaxy cluster.
The middle column shows the distribution of galaxies in the spectroscopic sample, selected in the same photometric redshift range. This selection provides a well-defined, complete sample of galaxies in and around the cluster, against which we compare our spectroscopic selection.
The third column presents the fraction between the two samples. 
Only bins with at least 10 galaxies are used.
Black density contours demonstrate that we may choose to ignore the noise around the edges.

The incompleteness plots consistently demonstrate that while there is a cut-off for faint, blue and small objects (i.e., in the x direction) in the spectroscopic sample, corrected for by applying weights to faint galaxies in our paper, it is not obvious that there are any trends with respect to other galaxy properties (i.e., in y-direction). 
The distribution of galaxies in the spectroscopic sample is compatible with that of the underlying photometric sample, which we consider to be complete. 
Since there are no obvious trends in any of the additionally probed parameters, we do not expect to be systematically incomplete in terms of color, size or compactness.

\end{appendices}

\end{document}